\documentclass[twoside,11pt]{article}

\usepackage{amsmath,amsfonts}
\usepackage{amssymb,latexsym}
\usepackage{amsxtra}
\usepackage{upref}
\usepackage{exscale}
\usepackage[mathscr]{eucal}

\newcommand{\be}{\begin{equation}}
\newcommand{\ee}{\end{equation}}
\newcommand{\beq}{\begin{eqnarray}}
\newcommand{\eeq}{\end{eqnarray}}

\newcommand{\bea}{\begin{array}}
\newcommand{\eea}{\end{array}}
\newcommand{\lb}{\label}
\newcommand{\mcal}{\mathcal}
\newcommand{\mscr}{\mathscr}
\newcommand{\mfrak}{\mathfrak}

\newcommand{\ts}{\textstyle}
\newcommand{\pp}{\partial}
\newcommand{\im}{\mathsf{i}}
\newcommand{\ppr}{^{\boldsymbol{\prime}}}

\newcommand{\pdag}{^{\dagger}}
\newcommand{\wt}{\widetilde}
\newcommand{\ovv}{\overline}
\newcommand{\TRAB}{\raisebox{-4pt}{$\mbox{Tr}\atop {\scriptstyle a,b}$}}

\newcommand{\scr}{\scriptstyle}
\newcommand{\scrscr}{\scriptscriptstyle}
\newcommand{\scz}{\scriptsize}

\newcommand{\msf}{\mathsf}
\newcommand{\Teta}[1]{\mscr{T}_{{\scrscr#1}}^{{\scrscr(\eta_{#1})}}}

\newcommand{\TT}{\mscr{T}}

\newcommand{\sdelta}{{\scr\Delta}}

\newcommand{\unityb}{\raisebox{-3pt}{\mbox{{\LARGE $\hat{1}$}}}}

\numberwithin{equation}{section}
\allowdisplaybreaks[4]

\begin{document}

\begin{center}
{\large\bf Monte-Carlo sampling of self-energy matrices within sigma-models} \vspace*{0.1cm}\\
{\large\bf derived from Hubbard-Stratonovich transformed coherent state path integrals} \vspace*{0.3cm}\\
{\bf Bernhard Mieck}\footnote{e-mail: "bjmeppstein@arcor.de"; freelance
activity; current location : Zum Kohlwaldfeld 16, 65817 Eppstein, Germany.} \vspace{0.3cm}\\  {\sf\small 
Extracted and enhanced version of 22-May-2012 from the presented poster at {\bf CERF 2011}, \vspace{0.3cm} \\
{\bf International Conference on {\large"Correlation Effects in Radiation Fields 2011"},} \vspace{0.3cm} \\ 
in {\bf Rostock, Sep. 12-16 2011}; }
\end{center}

\begin{abstract}
\noindent The 'Neumann-Ulam' Monte-Carlo sampling is described for the calculation of a matrix inversion or a Green function
in case of Hubbard-Stratonovich (HS-)transformed coherent state path integrals. We illustrate how to circumvent direct
numerical inversion of a matrix to its Green function by taking random walks of suitably chosen matrices within a path integral
of even- and complex-valued self-energy matrices. The application of a random walk sampling 
is given by the possible separation of the total matrix, e.g. that matrix which determines
the Green function from its inversion, into a part of unity minus (or plus) a matrix which only contains eigenvalues with absolute value 
smaller than one. This allows to expand the prevailing Green function around the unit matrix in a Taylor expansion with a
separated, special matrix of sufficiently small eigenvalues. The presented sampling method is particularly appropriate around the
saddle point solution of the self-energy in a sigma model by using random number generators. It is also capable
for random sampling of HS-transformed path integrals from fermionic fields which interact through
gauge invariant bosons according to Yang-Mills theories. \newline
\vspace*{0.1cm}

\vspace*{0.1cm}

\noindent 1st {\bf PACS} : 02.70.Ss , 02.70.Tt , 02.70.Uu ;\newline
\noindent 2nd{\bf PACS} : 05.10.Ln , 11.10.Lm , 11.10.-z .
\end{abstract}

\tableofcontents

\newpage

\section{Introduction} \lb{x1}

\subsection{Coherent state path integrals and the HS-transformation to self-energies} \lb{x11}

Coherent state path integrals with a quartic interaction of fields, especially those with a fermionic character, allow for the
transformation to even- and complex-valued, anomalous-doubled self-energy matrices \cite{Gil}-\cite{Per}. After such a Hubbard-Stratonovich
(HS-)transformation \cite{HS1,HS2}, one usually performs a coset decomposition of the total self-energy in combination with a saddle
point approximation\cite{Lern1}-\cite{pop2}. In most cases the self-energy matrix is restricted to the elements of the orthogonal \(\mbox{so}(\mfrak{N},\mfrak{N})\) or symplectic \(\mbox{sp}(\mfrak{N},\mfrak{N})\) Lie algebras for fermionic or bosonic degrees of freedom in the initial coherent state path integral, respectively. Therefore, the transformed, total self-energy matrix can be factorized into density-related and anomalous-doubled parts with a coset decomposition \(\mbox{SO}(\mfrak{N},\mfrak{N})\,/\,\mbox{U}(\mfrak{N})\otimes\mbox{U}(\mfrak{N})\) for fermions or 
\(\mbox{Sp}(\mfrak{N},\mfrak{N})\,/\,\mbox{U}(\mfrak{N})\otimes\mbox{U}(\mfrak{N})\) for bosons. This results into a gradient expansion
of the determinant with the coset matrices which precisely contain the 'Nambu'-doubled, coset matrix algebra field degrees of freedom
for a spontaneous symmetry breaking \cite{coh1,sig29}.

In this article we point to the remarkable property of coherent state path integrals concerning the 'Neumann-Ulam' Monte-Carlo sampling
of Green functions \cite{Ulam1}-\cite{Neg2}. The presented case avoids direct numerical computation of a matrix to its Green function, as e.g. by
recursive techniques for approximate inversion of partitioned sub-matrices (cf. Refs.\cite{Kramer1}-\cite{recur1}).
We describe the applicability of random walks within
suitably chosen matrices occurring in coherent state path integrals so that matrix inversion 
is obtained from random sampling of transition probabilities of matrix elements. This is combined with properly defined scores of residual weights.
The chosen system of Refs.\cite{pulse1,cerf1}(section 3.3)\footnote{Title: 'Fermionic coherent state path integral for ultra-short laser pulses
and transformation to a field theory of coset matrices'} is generic for other path integrals with self-energies as that of QCD so that lattice
QCD sampling with determinants can be replaced by the presented 'Neumann-Ulam' Monte-Carlo method which is also applicable to the
HS-transformed coherent state path integral of BCS-QCD in Ref.\cite{BCS-QCD,BCS-SM}\footnote{Title: 'BCS-like action and Lagrangian from the gradient
expansion of the determinant of Fermi fields in QCD-type, non-Abelian gauge theories with chiral anomalies'}. By way of example we
consider the model of Refs.\cite{pulse1,cerf1} (section 3.3) 
with a two-band, semiconductor-related solid in order to hint to the possible 'Neumann-Ulam' sampling.
The corresponding HS-transformation of the original fermionic coherent state path integral is achieved from dyadic products of anomalous-doubled,
anti-commuting fields which results into an even- and complex-valued self-energy matrix as an element of the \(\mbox{so}(\mfrak{N},\mfrak{N})\)
Lie algebra. The presented sampling method is particularly applicable around the saddle point solutions of the anomalous-doubled self-energy
so that the fluctuations can be determined around the saddle point approximation with random number generators. The implementation
of a random sampling for HS-transformed coherent state path integrals follows from the possible splitting of the total matrix, e.g. that matrix
which comprises the Green functions from its inversion, 
into a part of unity minus(or plus) a matrix which only consists of eigenvalues with absolute value
smaller than one. In consequence, the prevailing Green function can be expanded around unity in a Taylor expansion with the
separated matrix whose eigenvalues are sufficiently confined within the open interval between minus and plus one. 

\subsection{'Neumann-Ulam' matrix inversion} \lb{x12}

In the following we briefly review the 'Neumann-Ulam' matrix inversion in order to extend this sampling method to HS-transformed
coherent state path integrals, as e.g. given in Refs.\cite{pulse1,cerf1} and \cite{BCS-QCD,BCS-SM}. 
The 'Neumann-Ulam' Monte-Carlo sampling takes random
walks of a matrix \(\hat{M}_{ij}\) in order to achieve a matrix inversion \(\hat{M}_{ij}^{-1}\), as for a Green function \(\hat{G}_{ij}=\hat{M}_{ij}^{-1}\). However, this method can only be applied, provided that the matrix \(\hat{M}_{ij}\)
can be separated into a unit matrix \(\hat{1}_{ij}\) minus a matrix \(\hat{m}_{ij}\) whose eigenvalues are completely restricted
to the open interval \((-1,+1)\) so that following expansion holds
\begin{subequations}
\begin{align} \lb{x1_1a}
\hat{M}_{ij} &= \hat{1}_{ij} - \hat{m}_{ij}\;;\;\;(\mbox{eigenvalues of } \hat{m}_{ij}\in (-1,+1)\;)\;;  \\  \lb{x1_1b}
\hat{G}_{ij} &= \hat{M}_{ij}^{-1} = (\hat{1}-\hat{m})_{ij}^{-1} = \hat{1}_{ij} + \hat{m}_{ij} + \hat{m}_{ii_{2}}\,\hat{m}_{i_{2}j} +
\hat{m}_{ii_{2}}\,\hat{m}_{i_{2}i_{3}}\,\hat{m}_{i_{3}j} + \ldots \;\;\;.
\end{align}
\end{subequations}
The 'Neumann-Ulam' method uses a decomposition of the matrix \(\hat{m}_{ij}=\hat{\msf{p}}_{ij}\:\hat{\mathsf{w}}_{ij}\)
into transition probabilities \(\hat{\msf{p}}_{ij}>0\) and a fixed, residual weight \(\hat{\msf{w}}_{ij}\in\mathbb{C}\) so that
random walks within the matrix  \(\hat{m}_{ij}=\hat{\msf{p}}_{ij}\:\hat{\mathsf{w}}_{ij}\) of various order
\((\hat{m}^{k})_{ij}\) result into the matrix inversion of \(\hat{M}_{ij}^{-1}\)
\begin{subequations}
\begin{align}   \lb{x1_2a}
\hat{G}_{ij} &= \hat{M}_{ij}^{-1} = (\hat{1}-\hat{m})_{ij}^{-1} = \hat{1}_{ij} +
\langle\hat{\msf{p}}_{ij}\, \hat{\msf{w}}_{ij}\rangle + \langle\hat{\msf{p}}_{ii_{2}}\, \hat{\msf{w}}_{ii_{2}}\;\hat{\msf{p}}_{i_{2}j}\, \hat{\msf{w}}_{i_{2}j}\rangle  + \langle\hat{\msf{p}}_{ii_{2}}\, \hat{\msf{w}}_{ii_{2}}\;\hat{\msf{p}}_{i_{2}i_{3}}\, \hat{\msf{w}}_{i_{2}i_{3}}\;\hat{\msf{p}}_{i_{3}j}\, \hat{\msf{w}}_{i_{3}j}\rangle  + \ldots \;;  \\   \lb{x1_2b}
\hat{m}_{ij} &= \hat{\msf{p}}_{ij}\,\hat{\msf{w}}_{ij}\;\;\;.
\end{align}
\end{subequations}
In order to obtain a valid score, one has also to introduce a stopping probability \(\msf{P}\!_{i}=1-\sum_{j}\hat{\msf{p}}_{ij}\)
so that a particular random walk does not continue 'ad infinitum' within above series (\ref{x1_1b}) and (\ref{x1_2a}) !
A valid random walk \(\langle(\hat{m}^{k})_{ij}\rangle\) of order \(k\) within above summation of \(\hat{G}_{ij}=\hat{M}_{ij}^{-1}\)
has therefore to include the stopping probability \(\msf{P}\!_{i}\) into the 'total probability density \(\mscr{P}\)' of
transition probabilities and an additional factor \(1/\msf{P}\!_{i}\) into the 'score \(\mscr{S}\)' of the residual weights so that the
decomposition of \(\hat{m}_{ij} = \hat{\msf{p}}_{ij}\,\hat{\msf{w}}_{ij}\) still holds for a valid expansion of 
\( \hat{M}_{ij}^{-1} = (\hat{1}-\hat{m})_{ij}^{-1}\)
\begin{subequations}
\begin{align}\lb{x1_3a}
\langle(\hat{m}^{k})_{ij}\rangle & =\Big\langle \underbrace{\big(\hat{\msf{p}}_{ii_{2}}\,\hat{\msf{p}}_{i_{2}i_{3}}\cdots
\hat{\msf{p}}_{i_{k}i_{k+1}}\,\msf{P}\!_{i_{k+1}}\big)}_{\mbox{\scz 'probability density' } \mscr{P}} \times 
 \underbrace{\big(\hat{\msf{w}}_{ii_{2}}\,\hat{\msf{w}}_{i_{2}i_{3}}\cdots
\hat{\msf{w}}_{i_{k}i_{k+1}}\,\frac{1}{\msf{P}\!_{i_{k+1}}}\,\delta_{i_{k+1},j}\big)}_{\mbox{\scz 'score' } \mscr{S}_{ij}^{(k)}(\hat{\msf{w}}_{mn};{\scrscr\frac{1}{\msf{P}\!_{j}}})} \Big\rangle \;;  \\  \lb{x1_3b}
\msf{P}\!_{j} &=1-\sum_{i}\hat{\msf{p}}_{ji} \;\;\;.
\end{align}
\end{subequations}
As we compare the well-known sampling of an integral \(I=\int_{-\infty}^{+\infty}dx\;f(x)\) with a normalized probability density
\(\msf{p}(x)>0\) and residual weight function \(\msf{w}(x)=f(x)/p(x)\) to the presented case of random walks with a matrix \(\hat{m}_{ij}\)
\begin{subequations}
\begin{align}\lb{x1_4a}
I &= \int_{-\infty}^{+\infty}dx\;f(x) = \int_{-\infty}^{+\infty}dx\;\msf{p}(x)\;\msf{w}(x)  \;;  \\   \lb{x1_4b}
1 &=  \int_{-\infty}^{+\infty}dx\;\msf{p}(x)  \;;\;\;\;\msf{p}(x)>0\;;\;\;\;f(x)=\msf{p}(x)\;\msf{w}(x)\;\;,
\end{align}
\end{subequations}
one is guided to the proper definition of a probability measure in analogy to '\(dx\;\msf{p}(x)\)' and to the definition
of the corresponding score for every random walk of order \(k\). Note that the overall, total probability measure 
\(\mscr{P}\{\hat{\msf{p}},\msf{P}\}\) of random walks of all lengths is given by the sum of its sub-orders of definite length  \(m\). 
Every sampled sub-order \(\mscr{P}_{i_{1}\ldots i_{m+1}}^{(m)}\{\hat{\msf{p}}_{rs},\msf{P}\!_{i_{m+1}}\}\propto
\hat{\msf{p}}_{i_{1}i_{2}}\,\hat{\msf{p}}_{i_{2}i_{3}}\,\cdots\,\hat{\msf{p}}_{i_{m}i_{m+1}}\) ('\(m\)') selects
a particular sequence \(\hat{\msf{w}}_{i_{1}i_{2}}\,\hat{\msf{w}}_{i_{2}i_{3}}\,\cdots\,
\hat{\msf{w}}_{i_{m}i_{m+1}}\)  which is abbreviated by the 
specific score \(\mscr{S}_{ij}^{(k)}(\hat{\msf{w}}_{rs};\tfrac{1}{\msf{P}\!_{j}})\) with the same length '\(k=m\)'
and the same identical path of residual weights in between the start- and end-points \(i_{1}=i\), \(i_{m+1}=j\).
Since a random walk cannot continue for infinity, one has to insert the corrections of the stopping probability \(\msf{P}\!_{i_{m+1}}\)
and \(1/\msf{P}\!_{j}\) into probability measure and score, respectively
\begin{subequations}
\begin{align}\lb{x1_5a}
\mscr{P}\{\hat{\msf{p}},\msf{P}\} &= \sum_{m=0}^{\infty}
\mscr{P}_{i_{1}\ldots i_{m+1}}^{(m)}\{\hat{\msf{p}}_{rs},\msf{P}\!_{i_{m+1}}\}\;;\;\;\;\;
\mscr{P}_{i_{1} i_{1}}^{(m=0)}\{\hat{\msf{p}}_{rs},\msf{P}\!_{i_{1}}\} \equiv 1 \;;  \\   \lb{x1_5b}
\mscr{P}_{i_{1}\ldots i_{m+1}}^{(m)}\{\hat{\msf{p}}_{rs},\msf{P}\!_{i_{m+1}}\} &=
\hat{\msf{p}}_{i_{1}i_{2}}\,\hat{\msf{p}}_{i_{2}i_{3}}\,\cdots\,\hat{\msf{p}}_{i_{m}i_{m+1}}\;\msf{P}\!_{i_{m+1}} \;; \\  \lb{x1_5c} \mbox{\scz\sf assigned score : }\Longrightarrow\;
\mscr{S}_{ij}^{(k)}(\hat{\msf{w}}_{rs};\tfrac{1}{\msf{P}\!_{j}}) &= \delta_{i_{1},i}\;\hat{\msf{w}}_{i_{1}i_{2}}\,\hat{\msf{w}}_{i_{2}i_{3}}\,\cdots\,
\hat{\msf{w}}_{i_{k}i_{k+1}}\;\frac{1}{\msf{P}\!_{i_{k+1}}}\;\delta_{i_{k+1},j} \;;\;\;\;\;
\mscr{S}_{ij}^{(k=0)}(\hat{\msf{w}}_{rs};\tfrac{1}{\msf{P}\!_{j}}) \equiv 1_{ij} \;\;\;\; .
\end{align}
\end{subequations}
The 'Neumann-Ulam' averaging for the Green function \(\hat{G}_{ij}\), abbreviated by following brackets
\(\langle\ldots\rangle_{\scrscr N.U.}\), is therefore properly determined by using the probability measure
\(\mscr{P}_{i_{1}\ldots i_{k+1}}^{(k)}\{\hat{\msf{p}}_{rs},\msf{P}\!_{i_{k+1}}\}\) and its accompanying score
\(\mscr{S}_{ij}^{(k)}(\hat{\msf{w}}_{rs};\tfrac{1}{\msf{P}\!_{j}})\) for every order '\(l\)' of the expansion
\((\hat{1}-\hat{m})_{ij}^{-1}\)
\begin{align}\lb{x1_6}
&\hat{G}_{ij}  =\hat{M}_{ij}^{-1}=(\hat{1}-\hat{m})_{ij}^{-1}=
\Big\langle(\hat{1}-\hat{\msf{w}})_{ij}^{-1}\Big\rangle_{\scrscr N.U.}=
\hat{1}_{ij}+\sum_{k=1}^{\infty} \mscr{P}\{\hat{\msf{p}},\msf{P}\} \;\mscr{S}_{ij}^{(k)}(\hat{\msf{w}}_{rs};\tfrac{1}{\msf{P}\!_{j}}) =
\\   \notag &=\hat{1}_{ij} +  \sum_{k,m=1}^{\infty} \mscr{P}_{i_{1}\ldots i_{m+1}}^{(m)}\{\hat{\msf{p}}_{rs},\msf{P}\!_{i_{m+1}}\}\;
\delta_{m,k}\;\mscr{S}_{ij}^{(k)}(\hat{\msf{w}}_{rs};\tfrac{1}{\msf{P}\!_{j}}) =  \hat{1}_{ij} +
\sum_{m=1}^{\infty}\Big(\hat{\msf{p}}_{ii_{2}}\,\hat{\msf{p}}_{i_{2}i_{3}}\,\cdots\,\hat{\msf{p}}_{i_{m}j}\;\msf{P}\!_{j}\Big)\;
\Big(\hat{\msf{w}}_{ii_{2}}\,\hat{\msf{w}}_{i_{2}i_{3}}\,\cdots\,\hat{\msf{w}}_{i_{m}j}\;\frac{1}{\msf{P}\!_{j}}\Big)_{\!\!\mbox{.}}
\end{align}
Hence, the variance straightforwardly follows by application of the probability measure (\ref{x1_5a},\ref{x1_5b}) and 
the score (\ref{x1_5c}) and by introduction of an
additional matrix \(\hat{K}_{ij}=\hat{m}_{ij}\,\hat{\msf{w}}_{ij}\) which determines the deviation from the mean value combined
with the inverse of the stopping probability \(1/\msf{P}_{j}\). We emphasize that each score 
\(\mscr{S}_{ij}^{(k)}(\hat{\msf{w}}_{rs};\tfrac{1}{\msf{P}\!_{j}})\), \(\mscr{S}_{ij}^{(l)}(\hat{\msf{w}}_{rs};\tfrac{1}{\msf{P}\!_{j}})\)
of order '\(k\)' and '\(l\)' is specified by a sub-probability measure 
\(\mscr{P}_{i_{1}\ldots i_{k+1}}^{(k)}\{\hat{\msf{p}}_{rs},\msf{P}\!_{i_{k+1}}\}\),
\(\mscr{P}_{i_{1}\ldots i_{l+1}}^{(l)}\{\hat{\msf{p}}_{rs},\msf{P}\!_{i_{l+1}}\}\), respectively. As one takes a random walk of length
'\(m\)' according to the transition probabilities and stopping probability with the sub-probability measure 
\(\mscr{P}_{i_{1}\ldots i_{m+1}}^{(m)}\{\hat{\msf{p}}_{rs},\msf{P}\!_{i_{m+1}}\}\), one is also naturally confined to products of scores
\(\mscr{S}_{ij}^{(k)}(\hat{\msf{w}}_{rs};\tfrac{1}{\msf{P}\!_{j}})\), \(\mscr{S}_{ij}^{(l)}(\hat{\msf{w}}_{rs};\tfrac{1}{\msf{P}\!_{j}})\)
which have the same length '\(m\)' and the same identical path of transitions of the actual, sampled random walk with
\(\mscr{P}_{i_{1}\ldots i_{m+1}}^{(m)}\{\hat{\msf{p}}_{rs},\msf{P}\!_{i_{m+1}}\}\). In consequence the sub-probability measure 
\(\mscr{P}_{i_{1}\ldots i_{m+1}}^{(m)}\{\hat{\msf{p}}_{rs},\msf{P}\!_{i_{m+1}}\}\) always specifies the actual
random walk and restricts the allowed, possible scores 
\(\mscr{S}_{ij}^{(k)}(\hat{\msf{w}}_{rs};\tfrac{1}{\msf{P}\!_{j}})\), \(\mscr{S}_{ij}^{(l)}(\hat{\msf{w}}_{rs};\tfrac{1}{\msf{P}\!_{j}})\)
to the same identical steps and the length of the prevailing sub-probability measure. 
We have especially focused to this point because this leads to the important, two
Kronecker-deltas \(\delta_{m,k}\) and \(\delta_{m,l}\) in following expansion for the variance with
the sampled path \(\mscr{P}_{i_{1}\ldots i_{m+1}}^{(m)}\{\hat{\msf{p}}_{rs},\msf{P}\!_{i_{m+1}}\} =
\hat{\msf{p}}_{i_{1}i_{2}}\,\hat{\msf{p}}_{i_{2}i_{3}}\,\cdots\,\hat{\msf{p}}_{i_{m}i_{m+1}}\;\msf{P}\!_{i_{m+1}}\)
including the correction of the stopping probability
\begin{subequations}
\begin{align}\notag &
\big(\Delta\hat{G}_{ij}\big)^{2} = \Big\langle \Big((\hat{1}-\hat{\msf{w}})_{ij}^{-1}-
\langle(\hat{1}-\hat{\msf{w}})_{ij}^{-1}\rangle_{\scrscr N.U.}
\Big)^{2}\Big\rangle_{\scrscr N.U.}=\Big\langle(\hat{1}-\hat{\msf{w}})_{ij}^{-1}\;(\hat{1}-\hat{\msf{w}})_{ij}^{-1}\Big\rangle_{\scrscr N.U.} -
\Big(\Big\langle(\hat{1}-\hat{\msf{w}})_{ij}^{-1}\Big\rangle_{\scrscr N.U.}\Big)^{2}  =   \\  \lb{x1_7a} &=\sum_{k,l=0}^{\infty}
\Big\langle (\hat{\msf{w}}^{k})_{ij}\; (\hat{\msf{w}}^{l})_{ij}\Big\rangle_{\scrscr N.U.} -(\hat{G}_{ij})^{2}=\sum_{k,l=0}^{\infty}
\mscr{P}\{\hat{\msf{p}},\msf{P}\}\;\;
\mscr{S}_{ij}^{(k)}(\hat{\msf{w}}_{rs};\tfrac{1}{\msf{P}\!_{j}})\;\;\mscr{S}_{ij}^{(l)}(\hat{\msf{w}}_{rs};\tfrac{1}{\msf{P}\!_{j}})-
(\hat{G}_{ij})^{2} =     \\   \notag &=\hat{1}_{ij}+\sum_{k,l=1}^{\infty}\sum_{m=1}^{\infty}
\mscr{P}_{i_{1}\ldots i_{m+1}}^{(m)}\{\hat{\msf{p}}_{rs},\msf{P}\!_{i_{m+1}}\}\;\;\delta_{m,k}\;
\mscr{S}_{ij}^{(k)}(\hat{\msf{w}}_{rs};\tfrac{1}{\msf{P}\!_{j}})\;\;\delta_{m,l}\;
\mscr{S}_{ij}^{(l)}(\hat{\msf{w}}_{rs};\tfrac{1}{\msf{P}\!_{j}})-
(\hat{G}_{ij})^{2} =   \\   \notag  &= \hat{1}_{ij}+\sum_{m=1}^{\infty}
\mscr{P}_{i_{1}\ldots i_{m+1}}^{(m)}\{\hat{\msf{p}}_{rs},\msf{P}\!_{i_{m+1}}\}\;\delta_{i_{1},i}\;
(\hat{\msf{w}}_{ii_{2}}\,\hat{\msf{w}}_{ii_{2}})\cdot(\hat{\msf{w}}_{i_{2}i_{3}}\,\hat{\msf{w}}_{i_{2}i_{3}})\cdots
(\hat{\msf{w}}_{i_{m}i_{m+1}}\,\hat{\msf{w}}_{i_{m}i_{m+1}})\cdot\Big(\frac{1}{\msf{P}\!_{i_{m+1}}}\Big)^{2}
\delta_{i_{m+1},j} -(\hat{G}_{ij})^{2} =   \\  \notag &=\hat{1}_{ij}+\sum_{m=1}^{\infty}
\Big(\hat{\msf{p}}_{ii_{2}}\,\hat{\msf{p}}_{i_{2}i_{3}}\,\cdots\,\hat{\msf{p}}_{i_{m}j}\;\msf{P}\!_{j}\Big)
(\hat{\msf{w}}_{ii_{2}})^{2}\,(\hat{\msf{w}}_{i_{2}i_{3}})^{2}\,\cdots\,
(\hat{\msf{w}}_{i_{m}j})^{2}\Big(\frac{1}{\msf{P}\!_{j}}\Big)^{2}-(\hat{G}_{ij})^{2}  =  \\   \notag &= \hat{1}_{ij}+
\Big(\sum_{m=1}^{\infty}\hat{K}_{ii_{2}}\,\hat{K}_{i_{2}i_{3}}\,\cdots\,\hat{K}_{i_{m}j}\Big)\frac{1}{\msf{P}\!_{j}} 
-(\hat{G}_{ij})^{2}  =  \big(\hat{1}-\hat{K}\big)_{ij}^{-1}\;\;\frac{1}{\msf{P}\!_{j}}-(\hat{G}_{ij})^{2}  \;\;\;;  \\  \lb{x1_7b}
& \hat{K}_{ij} = \hat{\msf{p}}_{ij}\;(\hat{\msf{w}}_{ij})^{2}= \hat{m}_{ij}\;\hat{\msf{w}}_{ij}\;\;\;\;.
\end{align}
\end{subequations}
A variant of this sampling method is given by a partial removal of the stopping probability \(\msf{P}\!_{i}\).
If the random walk does not stop after '\(k\)' steps according to the conjugate probability \(\ovv{\msf{P}}\!_{i}=1-\msf{P}_{i}=
\sum_{j}\hat{\msf{p}}_{ij}\) for continuation, one simply adds the actual, obtained score 
\(\hat{\msf{w}}_{ii_{2}}\,\hat{\msf{w}}_{i_{2}i_{3}}\,\cdots\,\hat{\msf{w}}_{i_{k}j}\) {\it without} the inverted
stopping probability to the already existing, accumulated value of order '\(k\)' and generates another transition probability
\(\hat{\msf{p}}_{i_{k+1}i_{k+2}}\) for a further step '\(k+1\)'. One can thus proceed to higher order scores '\((\hat{\msf{w}}^{l})\)',
(\(l>k\)) with the actual random walk which only ceases as soon as a stopping probability \(\msf{P}\!_{l}\) is sampled in a higher order
'\(l>k\)' of the random walk. The corresponding probability measure and score has to be modified in order to take into account the partial 
removal of the stopping probability \(\msf{P}\!_{i}\) with the conjugate continuation probability  \(\ovv{\msf{P}}\!_{i}\).
We therefore distinguish between the two cases \(\msf{Q}_{i}=\msf{P}\!_{i}\) and \(\msf{Q}_{i}=\ovv{\msf{P}}\!_{i}\)
for stopping and continuing of the random walk in following modified definition of probability measure and score of a random walk
\begin{subequations}
\begin{align}\lb{x1_8a}
\msf{P}\!_{i} &= 1-\sum_{j}\hat{\msf{p}}_{ij} \;;  \quad\quad \ovv{\msf{P}}\!_{i}=1-\msf{P}_{i}= \sum_{j}\hat{\msf{p}}_{ij} \;;   \\  \lb{x1_8b}
\mscr{P}\{\hat{\msf{p}},\msf{P},\ovv{\msf{P}}\} &=\tfrac{1}{2}\big(\mscr{P}\{\hat{\msf{p}},\msf{P}\}+
\mscr{P}\{\hat{\msf{p}},\ovv{\msf{P}}\}\big)=\tfrac{1}{2}\sum_{\msf{Q}=\msf{P},\ovv{\msf{P}}}
\sum_{m=0}^{\infty}
\mscr{P}_{i_{1}\ldots i_{m+1}}^{(m)}\{\hat{\msf{p}}_{rs},\msf{Q}_{i_{m+1}}\}\;;\;\;\;\;
\mscr{P}_{i_{1} i_{1}}^{(m=0)}\{\hat{\msf{p}}_{rs},\msf{Q}_{i_{1}}\} \equiv 1 \;;  \\  \lb{x1_8c}
\mscr{P}_{i_{1}\ldots i_{m+1}}^{(m)}\{\hat{\msf{p}}_{rs},\msf{Q}_{i_{m+1}}\} &=
\hat{\msf{p}}_{i_{1}i_{2}}\,\hat{\msf{p}}_{i_{2}i_{3}}\,\cdots\,\hat{\msf{p}}_{i_{m}i_{m+1}}\;\msf{Q}_{i_{m+1}} \;; \quad
(\msf{Q}_{i_{m+1}}=\msf{P}\!_{i_{m+1}}\;\;\mbox{or }\msf{Q}_{i_{m+1}}=\ovv{\msf{P}}\!_{i_{m+1}})\;;  \\  \lb{x1_8d}
\mscr{S}_{ij}^{(k)}(\hat{\msf{w}}_{rs};\tfrac{1}{\msf{Q}_{j}}) &= \delta_{i_{1},i}\;\hat{\msf{w}}_{i_{1}i_{2}}\,\hat{\msf{w}}_{i_{2}i_{3}}\,\cdots\,
\hat{\msf{w}}_{i_{k}i_{k+1}}\;\frac{1}{\msf{Q}_{i_{k+1}}}\;\delta_{i_{k+1},j} \;;\;\;\;\;
\mscr{S}_{ij}^{(k=0)}(\hat{\msf{w}}_{rs};\tfrac{1}{\msf{Q}_{j}}) \equiv 1_{ij} \;\;\;\; .
\end{align}
\end{subequations}
According to relations (\ref{x1_5a}-\ref{x1_6}), one can define the modified 'Neumann-Ulam' sampling by the brackets
\(\langle\ldots\rangle_{\scrscr\wt{N.U.}}\) with the probability measure \(\mscr{P}\{\hat{\msf{p}},\msf{P},\ovv{\msf{P}}\} =\tfrac{1}{2}\big(\mscr{P}\{\hat{\msf{p}},\msf{P}\}+\mscr{P}\{\hat{\msf{p}},\ovv{\msf{P}}\}\big)\) (\ref{x1_8a}-\ref{x1_8c}) and score
\(\mscr{S}_{ij}^{(k)}(\hat{\msf{w}}_{rs};\tfrac{1}{\msf{Q}_{j}}) \) (\ref{x1_8d}) 
\begin{align}\lb{x1_9} &
\hat{G}_{ij}=\hat{M}_{ij}^{-1}=(\hat{1}-\hat{m})_{ij}^{-1}=
\Big\langle(\hat{1}-\hat{\msf{w}})_{ij}^{-1}\Big\rangle_{\scrscr\wt{N.U.}}=
\hat{1}_{ij}+\tfrac{1}{2}\sum_{\msf{Q}=\msf{P},\ovv{\msf{P}}}\sum_{k=1}^{\infty}
\mscr{P}\{\hat{\msf{p}},\msf{Q}\} \;\mscr{S}_{ij}^{(k)}(\hat{\msf{w}}_{rs};\tfrac{1}{\msf{Q}_{j}}) =
\\   \notag &=\hat{1}_{ij} + \tfrac{1}{2}\sum_{\msf{Q}=\msf{P},\ovv{\msf{P}}} \sum_{k,m=1}^{\infty} \mscr{P}_{i_{1}\ldots i_{m+1}}^{(m)}\{\hat{\msf{p}}_{rs},\msf{Q}_{i_{m+1}}\}\;
\delta_{m,k}\;\mscr{S}_{ij}^{(k)}(\hat{\msf{w}}_{rs};\tfrac{1}{\msf{Q}_{j}}) =  \\   \notag &= \hat{1}_{ij} + \tfrac{1}{2}\sum_{\msf{Q}=\msf{P},\ovv{\msf{P}}}
\sum_{m=1}^{\infty}\Big(\hat{\msf{p}}_{ii_{2}}\,\hat{\msf{p}}_{i_{2}i_{3}}\,\cdots\,\hat{\msf{p}}_{i_{m}j}\;\msf{Q}_{j}\Big)\;
\Big(\hat{\msf{w}}_{ii_{2}}\,\hat{\msf{w}}_{i_{2}i_{3}}\,\cdots\,\hat{\msf{w}}_{i_{m}j}\;\frac{1}{\msf{Q}_{j}}\Big) =  \\  \notag &=
\hat{1}_{ij}+\tfrac{1}{2}\sum_{m=1}^{\infty}\Big[\Big(\hat{\msf{p}}_{ii_{2}}\,\hat{\msf{p}}_{i_{2}i_{3}}\,\cdots\,
\hat{\msf{p}}_{i_{m}j}\;\msf{P}\!_{j}\Big)\;
\Big(\hat{\msf{w}}_{ii_{2}}\,\hat{\msf{w}}_{i_{2}i_{3}}\,\cdots\,\hat{\msf{w}}_{i_{m}j}\;\frac{1}{\msf{P}\!_{j}}\Big)
+  \\      \notag &\hspace*{1.0cm}+
\Big(\hat{\msf{p}}_{ii_{2}}\,\hat{\msf{p}}_{i_{2}i_{3}}\,\cdots\,\hat{\msf{p}}_{i_{m}j}\;(1-\msf{P}\!_{j})\Big)\;
\Big(\hat{\msf{w}}_{ii_{2}}\,\hat{\msf{w}}_{i_{2}i_{3}}\,\cdots\,\hat{\msf{w}}_{i_{m}j}\;\frac{1}{(1-\msf{P}\!_{j})}\Big)\Big]_{\mbox{.}}
\end{align}
In a similar manner we can conclude for the corresponding variance of the modified random walk 
\(\langle\ldots\rangle_{\scrscr\wt{N.U.}}\) with a partial removal of the stopping probability. As one applies the proper definitions
(\ref{x1_8a}-\ref{x1_9}) for the sampling, one achieves a variance with an additional term which is proportional
to the inverted continuation probability \(1/\ovv{\msf{P}}\!_{j}=1/(1-\msf{P}_{j})\) (cf. (\ref{x1_7a},\ref{x1_7b}))
\begin{align}\notag  &
\big(\Delta\hat{G}_{ij}\big)^{2} = \Big\langle \Big((\hat{1}-\hat{\msf{w}})_{ij}^{-1}-
\Big\langle(\hat{1}-\hat{\msf{w}})_{ij}^{-1}\Big\rangle_{\scrscr \wt{N.U.}}
\Big)^{2}\Big\rangle_{\scrscr \wt{N.U.}}=\Big\langle(\hat{1}-\hat{\msf{w}})_{ij}^{-1}\;(\hat{1}-\hat{\msf{w}})_{ij}^{-1}
\Big\rangle_{\scrscr \wt{N.U.}} -
\Big(\Big\langle(\hat{1}-\hat{\msf{w}})_{ij}^{-1}\Big\rangle_{\scrscr \wt{N.U.}}\Big)^{2}  = \\    \lb{x1_10}  &=  \big(\hat{1}-\hat{K}\big)_{ij}^{-1}\;\;\frac{1}{2}\Big(\frac{1}{\msf{P}\!_{j}}+\frac{1}{(1-\msf{P}\!_{j})}\Big)-(\hat{G}_{ij})^{2}  \;;\quad
 \hat{K}_{ij} = \hat{\msf{p}}_{ij}\;(\hat{\msf{w}}_{ij})^{2}= \hat{m}_{ij}\;\hat{\msf{w}}_{ij}\;\;\;\;.
\end{align}
One might object that the possible split \(\hat{M}_{ij}=\hat{1}_{ij}-\hat{m}_{ij}\) already simplifies the calculation of the
corresponding Green function \(\hat{G}_{ij}=\hat{M}_{ij}^{-1}=(\hat{1}-\hat{m})_{ij}^{-1}=\hat{1}_{ij}+\hat{m}_{ii_{2}}\hat{m}_{i_{2}j}+\ldots\)
in such a manner that it is sufficient to compute matrix multiplications \((\hat{m}^{k})_{ij}\) at varios powers '\(k\)'
without any Monte-Carlo sampling or random walks of transition probabilities. This may be advantageous for strictly banded, 1-dim.\  cases;
however, as more and more matrix multiplications are performed to higher orders \((\hat{m}^{k})_{ij}\), the spread of an initially
sparse or banded matrix \(\hat{m}_{ij}\) immediately leads to rather occupied, full matrices which tend to the full amount
of multiplications \(N_{\hat{m}}^{3}\) \((\hat{m}_{ij}; i,j=1,\ldots,N_{\hat{m}})\) at every order '\(k\)' in the Taylor expansion.
Therefore, the matrix multiplications of the Taylor expansion up to maximal power '\(k_{max}\)' without random sampling will
finally approach the number '\(\sim k_{max}\,N_{\hat{m}}^{3}\)' of floating point operations even for an initially sparse matrix.
It may then even be more adequate to determine the full matrix inversion \(\hat{G}_{ij}=(\hat{1}-\hat{m})_{ij}^{-1}\)
without any Taylor expansion by 'direct' numerical inversion which also is of the order of '\(\sim N_{\hat{m}}^{3}\)'
floating point operations. Hence, we can point out the advantage of the presented Monte-Carlo sampling for 'Neumann-Ulam'
matrix inversion of Green functions; the advantage of a vanishing spread of the initial (sparsely distributed) matrix entries of  \(\hat{m}_{ij}\)
is kept throughout all random walks as higher orders of the Taylor expansion
are considered up to the end of the individual random walk with stopping probability \(\msf{P}\!_{j}\). The reliability and
effectiveness of the described Monte-Carlo sampling for a matrix inversion finally depends on the statistical error which is determined
by the matrix \(\hat{K}_{ij} = \hat{m}_{ij}\;\hat{\msf{w}}_{ij}\) and the stopping and continuation probabilities
\(\msf{P}\!_{j},\;\ovv{\msf{P}}\!_{j}=1-\msf{P}\!_{j}\) in the variances (\ref{x1_7a},\ref{x1_7b}) and (\ref{x1_10}), respectively.

\section{Random walks applied to HS-transformed coherent state path integrals} \lb{x2}

\subsection{Self-energy $\hat{\Sigma}_{\mu_{1},s_{1};\mu_{2},s_{2}}^{ab}(\Teta{j_{1}}\!,\vec{x}_{1};\Teta{j_{2}}\!,\vec{x}_{2})$
of densities '$a=b$' and of 'Nambu'-parts '$a\neq b$'} \lb{x21}

In Ref.\cite{pulse1} we have outlined the derivation of the path integral in terms of the anomalous-doubled self-energy which one attains after
a HS-transformation of the dyadic product of 'Nambu'-doubled, anti-commuting fields. In this section we start out from the final result
of section 3.3 in Ref.\cite{pulse1} and introduce additional abbreviations for the contour time steps \(\Teta{j}\), the spatial coordinate vector \(\vec{x}\),
the electron-, hole- (\(\mu=e,h\)) and spin-labels (\(s=\uparrow,\downarrow\)) of the semiconductor-related solid. Furthermore, we
have to include 'Nambu'-indices '\(a,b,c,\ldots=1,2\)' for the anomalous-doubling of the fermionic fields where the block-diagonal parts
'\(11\)', '\(22\)' and off-diagonal parts '\(12\)', '\(21\)' of the self-energy matrix 
\(\hat{\Sigma}_{\mu_{1},s_{1};\mu_{2},s_{2}}^{ab}(\Teta{j_{1}}\!,\vec{x}_{1};\Teta{j_{2}}\!,\vec{x}_{2})\)
specify the density-related and anomalous-doubled parts, respectively. We extremely simplify notations in order to display
the structure of the path integral with the self-energy matrix as an element of the \(\mbox{so}(\mfrak{N},\mfrak{N})\) Lie algebra.
The overall dimension '\(\mfrak{N}\)' of latter Lie algebra \(\mbox{so}(\mfrak{N},\mfrak{N})\) is determined by the product of the
respective two degrees of the band-label  (\(\mu=e,h\)), the spin-label (\(s=\uparrow,\downarrow\)) as well as the time contour index
(\(\eta_{j}=\pm\)) multiplied by the number of chosen discrete time steps \((N=t_{N}/\Delta t)\)
and the number \((\mcal{N}_{x}=(L/\Delta x)^{d})\) of spatial points on an underlying, '\(d\)'-dimensional grid of system length '\(L\)'
with space intervals \(\Delta x\)
\begin{align}\lb{x2_1}
\mbox{dimension }\mfrak{N}\hspace*{0.2cm} \mbox{of } \mbox{so}(\mfrak{N},\mfrak{N})&=\underbrace{(\mu=\mbox{'e','h'})}_{2}\times
\underbrace{(s=\uparrow,\downarrow)}_{2}\times\underbrace{(\eta_{j}=\pm)}_{2}\times\underbrace{(N=t_{N}/\Delta t)}_{N}\times
\underbrace{(\mcal{N}_{x}=(L/\Delta x)^{d})}_{\mcal{N}_{x}}\;\;\;.
\end{align}
We abbreviate the set of variables 'contour time \(\Teta{j}\)', 'spatial coordinate vector \(\vec{x}\)', 'band-label \((\mu=e,h)\)'
and spin index '\((s=\uparrow,\downarrow)\)' by the uppercase letter '\(\mfrak{I}\)' and the restriction to 'contour time \(\Teta{j}\)'
and 'spatial vector \(\vec{x}\)' by the lowercase letter \(\mfrak{i}\)
\begin{align}\lb{x2_2}
(\Teta{j}\!\!,\vec{x};\mu,s)&=\mfrak{I}\;;\quad (\Teta{j}\!\!,\vec{x})=\mfrak{i}\;.
\end{align}
The notation of the self-energy matrix \(\hat{\Sigma}_{\mu_{1},s_{1};\mu_{2},s_{2}}^{ab}(\Teta{j_{1}}\!,\vec{x}_{1};\Teta{j_{2}}\!,\vec{x}_{2})\)
is thus transformed to \(\hat{\Sigma}_{\mfrak{I}_{1}\mfrak{I}_{2}}^{ab}\)
\begin{align}\lb{x2_3}
\hat{\Sigma}_{\mu_{1},s_{1};\mu_{2},s_{2}}^{ab}(\Teta{j_{1}}\!,\vec{x}_{1};\Teta{j_{2}}\!,\vec{x}_{2})\rightarrow
\hat{\Sigma}_{\mfrak{I}_{1}\mfrak{I}_{2}}^{ab}\;,
\end{align}
where additionally attached subscripts '\(1\)', '\(2\)' of '\(\mfrak{I}_{1}\)', '\(\mfrak{I}_{2}\)' (or respectively the reduced case
 '\(\mfrak{i}_{1}\)', '\(\mfrak{i}_{2}\)' ) distinguish between the two sets
of contour-time-coordinate labels including band- and spin indices for the uppercase letters (\ref{x2_4a}) 
or respectively lowercase letters (\ref{x2_4b})
\begin{subequations}
\begin{align}\lb{x2_4a}
\mfrak{I}_{1} &=(\Teta{j_{1}}\!\!,\vec{x}_{1};\mu_{1},s_{1})\;;\;
\mfrak{I}_{2} =(\Teta{j_{2}}\!\!,\vec{x}_{2};\mu_{2},s_{2})\;;\;\ldots\;;    \\     \lb{x2_4b}
\mfrak{i}_{1} &=(\Teta{j_{1}}\!\!,\vec{x}_{1})\;;\;\mfrak{i}_{2}=(\Teta{j_{2}}\!\!,\vec{x}_{2})\;;\;\ldots \;.
\end{align}
\end{subequations}
According to the dyadic product of anomalous-doubled, fermionic fields, the orthogonal symmetry \(\mbox{so}(\mfrak{N},\mfrak{N})\)
causes 'Nambu'-parts (\(b\neq a\)) \cite{Nambu,Gold} of an overall, hermitian self-energy matrix
\(\hat{\Sigma}_{\mfrak{I}_{1}\mfrak{I}_{2}}^{ab}\)
whose diagonal block parts are related by opposite sign and transposition and whose
two, anti-symmetric 'Nambu'-parts are related by hermitian conjugation
\begin{align}   \lb{x2_5}  \bea{rclrclrcl}
\hat{\Sigma}_{\mfrak{I}_{1}\mfrak{I}_{2}}^{ab}\hspace*{-0.36cm}
 &=&\hspace*{-0.36cm} \bigg(\bea{cc}
\hat{\Sigma}_{\mfrak{I}_{1}\mfrak{I}_{2}}^{11} &
\hat{\Sigma}_{\mfrak{I}_{1}\mfrak{I}_{2}}^{12}  \\
\hat{\Sigma}_{\mfrak{I}_{1}\mfrak{I}_{2}}^{21}  &
\hat{\Sigma}_{\mfrak{I}_{1}\mfrak{I}_{2}}^{22}
 \eea\bigg)^{ab}; &
\hat{\Sigma}_{\mfrak{I}_{1}\mfrak{I}_{2}}^{aa}\hspace*{-0.36cm}
 &=&\hspace*{-0.36cm}  \bigl(
\hat{\Sigma}_{\mfrak{I}_{1}\mfrak{I}_{2}}^{aa}\bigr)\pdag; &
\hat{\Sigma}_{\mfrak{I}_{1}\mfrak{I}_{2}}^{22} \hspace*{-0.36cm}
 &=&\hspace*{-0.36cm}  -\bigl(
\hat{\Sigma}_{\mfrak{I}_{1}\mfrak{I}_{2}}^{11}\bigr)^{T} ;   \\
\hat{\Sigma}_{\mfrak{I}_{1}\mfrak{I}_{2}}^{21} \hspace*{-0.36cm}
 &=&\hspace*{-0.36cm}  \bigl(
\hat{\Sigma}_{\mfrak{I}_{1}\mfrak{I}_{2}}^{12}\bigr)\pdag ; &
\hat{\Sigma}_{\mfrak{I}_{1}\mfrak{I}_{2}}^{(a\neq b)} \hspace*{-0.36cm}
 &=&\hspace*{-0.36cm}  -\bigl(
\hat{\Sigma}_{\mfrak{I}_{1}\mfrak{I}_{2}}^{(a\neq b)}\bigr)^{T}; &
\hat{\Sigma}_{\mfrak{I}_{1}\mfrak{I}_{2}}^{ab} \hspace*{-0.36cm}
 &\equiv&\hspace*{-0.36cm}  0\quad (\mbox{for}(j_{1}\mbox{\scz 'or'} j_{2}=0)\;\mbox{\scz 'or'}\;
(j_{1}\mbox{\scz 'or'} j_{2}=2N+1)\,).
\eea
\end{align}
A term as 
\begin{align}\lb{x2_6}
\sum_{'\!1','\!2'}
\TRAB\Big[(\hat{\Sigma}_{\mfrak{I}_{1}\mfrak{I}_{2}}^{ab}/\mcal{N}_{x})\cdot
(\hat{\Sigma}_{\mfrak{I}_{2}\mfrak{I}_{1}}^{ba}/\mcal{N}_{x})\Big]\quad,
\end{align}
with the distinguishing subscripts  '\(1\)', '\(2\)' of '\(\mfrak{I}_{1}\)', '\(\mfrak{I}_{2}\)' (\ref{x2_4a})
therefore implies a summation of the contour times '\(\Teta{j_{1}}\)',  '\(\Teta{j_{2}}\)', the spatial
coordinate vectors '\(\vec{x}_{1}\)', '\(\vec{x}_{2}\)', the band- and spin-labels '\(\mu_{1},\mu_{2}\)', '\(s_{1},s_{2}\)'
apart from the supplementary trace over 'Nambu'-space \((a,b=1,2)\) with the anomalous-doubled components of the self-energy.
Note, that we use normalized, spatial summations with factor \(1/\mcal{N}_{x}\) which we can absorb by a rescaling
of the self-energy within our derived path integral of section 3.3 in Ref.\cite{pulse1}. (The parameter \(\mcal{N}_{x}\) of the total number
of spatial points is similar to the parameter of the matrix dimensions in random matrix theories \cite{loggas,Pastur1}.)
\begin{align}\lb{x2_7}
\hat{\Sigma}_{\mfrak{I}_{1}\mfrak{I}_{2}}^{ab}/\mcal{N}_{x}\rightarrow\hat{\Sigma}_{\mfrak{I}_{1}\mfrak{I}_{2}}^{ab}\;.
\end{align}
Aside from the sets '\(\mfrak{I}_{1}\)', '\(\mfrak{I}_{2}\)', '\(\ldots\)', '\(\mfrak{i}_{1}\)', '\(\mfrak{i}_{2}\)', '\(\ldots\)' (\ref{x2_4a},\ref{x2_4b}) of
collected variables, we abbreviate the charge index '\(q_{\mu_{1}}=\pm 1 \)', '\(q_{\mu_{2}}=\pm 1 \)' of electrons 
\((\mu_{1},\mu_{2}=e)\) and holes \((\mu_{1},\mu_{2}=h)\) and the
contour time index '\(\eta_{j_{1}}=\pm \)', '\(\eta_{j_{2}}=\pm \)' in a similar manner to relations (\ref{x2_2},\ref{x2_3}) by subscripts  '\(1\)', '\(2\)' 
\begin{subequations}
\begin{align}\lb{x2_8a}
q_{1} &= q_{\mu_{1}}\;;\;q_{2} = q_{\mu_{2}}\;;\;\ldots \;;  \\   \lb{x2_8b}
\eta_{1} &= \eta_{j_{1}}\;;\;\eta_{2} = \eta_{j_{2}}\;;\;\ldots\;,
\end{align}
\end{subequations}
in the overall summations of the contour-time and space coordinates with the band- and spin-degrees of freedom. Moreover,
one has to regard the 'Nambu'-metric tensors \(\hat{\mathrm{I}}^{ab}\), \(\hat{\mathrm{S}}^{ab}\) which have to be inserted for 
the proper dyadic product of fermionic fields as the HS-transformation is performed for the remaining self-energy
\(\hat{\Sigma}_{\mfrak{I}_{1}\mfrak{I}_{2}}^{ab}\) of the \(\mbox{so}(\mfrak{N},\mfrak{N})\) Lie algebra
\begin{align}\lb{x2_9}
\hat{\mathrm{I}}^{ab}  &=\delta_{ab}\;\;\text{diag}\bigl\{\underbrace{\hat{1}}_{a=1}\;\boldsymbol{,}\;
\underbrace{\hat{\im}}_{a=2}\bigr\}\;;  \hspace*{0.3cm}
\hat{\mathrm{S}}^{ab} =\delta_{ab}\;\;
\text{diag}\bigl\{\underbrace{\hat{1}}_{a=1}\;\boldsymbol{,}\;\underbrace{-\hat{1}}_{a=2}\bigr\} =
\hat{\mathrm{I}}^{ac}\cdot\hat{\mathrm{I}}^{cb};(a,b,c= 1,2)\;.
\end{align}
After a rescaling of the one-particle operator \(\hat{\mathsf{H}}_{\mfrak{I}_{1}\mfrak{I}_{2}}^{ab}\) and the source term 
\(\hat{\mscr{J}}_{\mfrak{I}_{1}\mfrak{I}_{2}}^{ab}\) as for the self-energy \(\hat{\Sigma}_{\mfrak{I}_{1}\mfrak{I}_{2}}^{ab}\)
\begin{align}\lb{x2_10}
 \hat{\mathsf{H}}_{\mfrak{I}_{1}\mfrak{I}_{2}}^{ab}/\mcal{N}_{x}
 \Longrightarrow\hat{\mathsf{H}}_{\mfrak{I}_{1}\mfrak{I}_{2}}^{ab}\;;\;\;\;
\hat{\mscr{J}}_{\mfrak{I}_{1}\mfrak{I}_{2}}^{ab}/\mcal{N}_{x}
\Longrightarrow\hat{\mscr{J}}_{\mfrak{I}_{1}\mfrak{I}_{2}}^{ab}\;;
\end{align}
the ensemble-averaged generating \(\ovv{Z[\hat{\mscr{J}}]}\) of section 3.3 in Ref.\cite{pulse1} can be specified 
in terms of a notation which signifies a similarity to random matrix theories
\begin{align} \lb{x2_11}   &
\ovv{Z[\hat{\mscr{J}}]} =\int d[\hat{\Sigma}_{\mfrak{I}_{2}\mfrak{I}_{1}}^{ba}]\;
\exp\bigg\{-\tfrac{\im}{4}\tfrac{\sdelta t}{\hbar}\sum_{'\!1','\!2'}(\hat{\mscr{V}}_{\mfrak{i}_{2}\mfrak{i}_{1}})^{-1}
\TRAB\Big[\hat{\mathrm{S}}^{bb}\;q_{2}\;\eta_{2}\;\hat{\Sigma}_{\mfrak{I}_{2}\mfrak{I}_{1}}^{ba}\;
\hat{\mathrm{S}}^{aa}\;q_{1}\;\eta_{1}\;\hat{\Sigma}_{\mfrak{I}_{1}\mfrak{I}_{2}}^{ab}\Big]\bigg\}\times  \\ \notag &\times
\mbox{DET}\!\Big(\Big[
\hat{\mathrm{I}}^{bb}\;\hat{\mathsf{H}}_{\mfrak{I}_{2}\mfrak{I}_{1}}^{ba}\;\hat{\mathrm{I}}^{aa}+
\eta_{2}\;\hat{\mscr{J}}_{\mfrak{I}_{2}\mfrak{I}_{1}}\;\eta_{1}+
\hat{\mathrm{S}}^{bb}\;q_{2}\;\eta_{2}\;\Big(\im\tfrac{\sdelta t}{\hbar}\Big)\,\hat{\Sigma}_{\mfrak{I}_{2}\mfrak{I}_{1}}^{ba}\;
\hat{\mathrm{S}}^{aa}\;q_{1}\;\eta_{1}\Big]_{\mfrak{I}_{2}\mfrak{I}_{1}}^{ba}\Big)^{\boldsymbol{+\tfrac{1}{2}}}\;.
\end{align}
Sections 2.1 to 3.3 in Ref.\cite{pulse1}  consist of the derivation with the HS-transformation from a path integral
of anti-commuting fields for a semiconductor-related solid coupled to an ultra-short laser pulse with an inter-band
dipole moment. The latter, driving light-semiconductor interaction causes off-diagonal '\( 12\)', '\(21\)' 'Nambu'-blocks
in the one-particle operator  \(\hat{\mathsf{H}}_{\mfrak{I}_{1}\mfrak{I}_{2}}^{(a\neq b)}\) 
apart from the usual density-related kinetic energy terms \(\hat{\mathsf{H}}_{\mfrak{I}_{1}\mfrak{I}_{2}}^{(a=b)}\)
with further possible one-particle potentials along the block-diagonal parts '\( 11\)', '\(22\)'.
The Coulomb interaction \(V(\vec{x}_{1},\vec{x}_{2})\) and the
correlation function \(f(\vec{x}_{1},t_{1};\vec{x}_{2},t_{2})\) of disorder and noise are combined into the generalized interaction
potential \(\hat{\mscr{V}}_{\mfrak{i}_{2}\mfrak{i}_{1}}=\hat{\mscr{V}}(\vec{x}_{2},\Teta{j_{2}};\vec{x}_{1},\Teta{j_{1}})\)
\begin{align} \lb{x2_12} &
\hat{\mscr{V}}_{\mfrak{i}_{2}\mfrak{i}_{1}} = \eta_{j_{1}}\;\delta_{j_{2},j_{1}}\;
\bigl(\delta_{\eta_{{j}_{2}},\eta_{{j}_{1}}}\bigr)\;
V(\vec{x}_{2},\vec{x}_{1})\;-\;
\tfrac{\im}{2}\;\tfrac{\sdelta t}{\hbar}\;\;u_{0}^{2}\;\;
f\bigl(\vec{x}_{2},\TT_{j_{2}};\vec{x}_{1},\TT_{j_{1}}\bigr); \;\;(j_{1},j_{2}=1,\ldots,2N)\;,
\end{align}
which enters by the quotient \((\hat{\mscr{V}}_{\mfrak{i}_{2}\mfrak{i}_{1}})^{-1}\) 
for the quartic interaction of Fermi fields into the Gaussian factor with the two
self-energies \(\hat{\Sigma}_{\mfrak{I}_{2}\mfrak{I}_{1}}^{ba}\), \(\hat{\Sigma}_{\mfrak{I}_{1}\mfrak{I}_{2}}^{ab}\).
The latter matrices are accompanied by the charge- '\(q_{1},\,q_{2}\)' and contour-time '\(\eta_{1},\,\eta_{2}\)' indices including the diagonal
'Nambu'-metric tensors \(\hat{\mathrm{S}}^{aa}\), \(\hat{\mathrm{S}}^{bb}\) .
As we apply the simplified notations (\ref{x2_2}-\ref{x2_8b}) and the rescaling of the self-energy matrix \(\hat{\Sigma}_{\mfrak{I}_{1}\mfrak{I}_{2}}^{ab}\) (\ref{x2_7}), of the 
the one-particle operator \(\hat{\mathsf{H}}_{\mfrak{I}_{2}\mfrak{I}_{1}}^{ba}\) and of the source term
\(\hat{\mscr{J}}_{\mfrak{I}_{2}\mfrak{I}_{1}}\) (\ref{x2_10})
to the derived path integral (\ref{x2_11}) from section 3.3 of Ref.\cite{pulse1}  , 
one obtains following expression \(\ovv{Z[\hat{\mscr{J}}]}\) (\ref{x2_13}) of the generating
function whose 'exponential trace-log' relation replaces the determinant in (\ref{x2_11}). The pre-factor \(\tfrac{1}{2}\)
is caused by the square-root operation of the determinant in (\ref{x2_11}) with the additional sum over the discrete contour time steps
beginning at \(j=0\) and ending at \(j=2N+1\). We have separated the total number of spatial points \(\mcal{N}_{x}\)
as a parameter for the saddle point approximation by using a normalized spatial sum over the discrete grid points
\begin{align} \lb{x2_13}  &
\ovv{Z[\hat{\mscr{J}}]} =\int d[\hat{\Sigma}_{\mfrak{I}_{2}\mfrak{I}_{1}}^{ba}]\;
\exp\bigg\{-\tfrac{\im}{4}\tfrac{\sdelta t}{\hbar}\sum_{'\!1','\!2'}(\hat{\mscr{V}}_{\mfrak{i}_{2}\mfrak{i}_{1}})^{-1}
\TRAB\Big[\hat{\mathrm{S}}^{bb}\;q_{2}\;\eta_{2}\;\hat{\Sigma}_{\mfrak{I}_{2}\mfrak{I}_{1}}^{ba}\;
\hat{\mathrm{S}}^{aa}\;q_{1}\;\eta_{1}\;\hat{\Sigma}_{\mfrak{I}_{1}\mfrak{I}_{2}}^{ab}\Big]\bigg\}\times  \\ \notag &\times
\exp\bigg\{\tfrac{1}{2}\mcal{N}_{x}\sum_{\vec{x}}\sum_{j=0}^{2N+1}\TRAB\ln\Big[
\hat{\mathrm{I}}^{bb}\;\hat{\mathsf{H}}_{\mfrak{I}_{2}\mfrak{I}_{1}}^{ba}\;\hat{\mathrm{I}}^{aa}+
\eta_{2}\;\hat{\mscr{J}}_{\mfrak{I}_{2}\;\mfrak{I}_{1}}\eta_{1}+
\hat{\mathrm{S}}^{bb}\;q_{2}\;\eta_{2}\;\Big(\im\tfrac{\sdelta t}{\hbar}\Big)\,\hat{\Sigma}_{\mfrak{I}_{2}\mfrak{I}_{1}}^{ba}\;
\hat{\mathrm{S}}^{aa}\;q_{1}\;\eta_{1}\Big]\bigg\}\;\;;    \\   \notag &
\sum_{\vec{x}}\ldots \;=\frac{1}{\mcal{N}_{x}}\times\Big[\mbox{sum over the discrete spatial grid points}\Big]\;\ldots  \;\;\;.
\end{align}
The familiar kind of the generating function \(\ovv{Z[\hat{\mscr{J}}]}\) (\ref{x2_11},\ref{x2_13},\ref{x2_14a}) for a sigma model gives rise to a
saddle point equation (\ref{x2_14b}) with solution \(\hat{\Sigma}_{\mfrak{I}_{2}\mfrak{I}_{1}}^{\boldsymbol{(0)}ba}\) from a first order variation 
\(\sdelta\hat{\Sigma}_{\mfrak{I}_{2}\mfrak{I}_{1}}^{ba}\) (\ref{x2_14c}) with respect to the original
self-energy \(\hat{\Sigma}_{\mfrak{I}_{2}\mfrak{I}_{1}}^{ba}\)
\begin{subequations}
\begin{align} \notag 
&\ovv{Z[\hat{\mscr{J}}]} =\int d[\hat{\Sigma}_{\mfrak{I}_{2}\mfrak{I}_{1}}^{ba}]\;
\exp\bigg\{-\tfrac{\im}{4}\tfrac{\sdelta t}{\hbar}\sum_{'\!1','\!2'}(\hat{\mscr{V}}_{\mfrak{i}_{2}\mfrak{i}_{1}})^{-1}
\TRAB\Big[\hat{\mathrm{S}}^{bb}\;q_{2}\;\eta_{2}\;\Big(\hat{\Sigma}_{\mfrak{I}_{2}\mfrak{I}_{1}}^{\boldsymbol{(0)}ba} -
\sdelta\hat{\Sigma}_{\mfrak{I}_{2}\mfrak{I}_{1}}^{ba}\Big)\;
\hat{\mathrm{S}}^{aa}\;q_{1}\;\eta_{1}\;\Big(\hat{\Sigma}_{\mfrak{I}_{1}\mfrak{I}_{2}}^{\boldsymbol{(0)}ab} -
\sdelta\hat{\Sigma}_{\mfrak{I}_{1}\mfrak{I}_{2}}^{ab}\Big)\Big]\bigg\}\times  \\  \lb{x2_14a}  &\times
\exp\bigg\{\tfrac{1}{2}\mcal{N}_{x}\sum_{\vec{x}}\sum_{j=0}^{2N+1}\TRAB\ln\Big[
\hat{\mathrm{I}}^{bb}\;\hat{\mathsf{H}}_{\mfrak{I}_{2}\mfrak{I}_{1}}^{ba}\;\hat{\mathrm{I}}^{aa}+
\eta_{2}\;\hat{\mscr{J}}_{\mfrak{I}_{2}\mfrak{I}_{1}}\;\eta_{1}+
\hat{\mathrm{S}}^{bb}q_{2}\eta_{2}\Big(\im\tfrac{\sdelta t}{\hbar}\Big)
\Big(\hat{\Sigma}_{\mfrak{I}_{2}\mfrak{I}_{1}}^{\boldsymbol{(0)}ba} -\sdelta\hat{\Sigma}_{\mfrak{I}_{2}\mfrak{I}_{1}}^{ba}\Big)
\hat{\mathrm{S}}^{aa}q_{1}\eta_{1}\Big]\bigg\};         \\     \lb{x2_14b}&
(\hat{\mscr{V}}_{\mfrak{i}_{2}\mfrak{i}_{1}})^{-1}\;\hat{\Sigma}_{\mfrak{I}_{2}\mfrak{I}_{1}}^{\boldsymbol{(0)}ba} =
\hat{\mscr{G}}_{\mfrak{I}_{2}\mfrak{I}_{1}}^{\boldsymbol{(0)}ba}=
\Big[\hat{\mathrm{I}}^{b\ppr b\ppr}\;\hat{\mathsf{H}}_{\mfrak{I}_{4}\mfrak{I}_{3}}^{b\ppr a\ppr}\;\hat{\mathrm{I}}^{a\ppr a\ppr}+
\eta_{4}\;\hat{\mscr{J}}_{\mfrak{I}_{4}\mfrak{I}_{3}}\;\eta_{3}+
\hat{\mathrm{S}}^{b\ppr b\ppr}\;q_{4}\;\eta_{4}\;
\Big(\im\tfrac{\sdelta t}{\hbar}\Big)\,\hat{\Sigma}_{\mfrak{I}_{4}\mfrak{I}_{3}}^{\boldsymbol{(0)}b\ppr a\ppr}\;
\hat{\mathrm{S}}^{a\ppr a\ppr}\;q_{3}\;\eta_{3}\Big]_{\mfrak{I}_{2}\mfrak{I}_{1}}^{\boldsymbol{-1};ba};   \\   \lb{x2_14c}
&\hat{\Sigma}_{\mfrak{I}_{2}\mfrak{I}_{1}}^{ba} = \hat{\Sigma}_{\mfrak{I}_{2}\mfrak{I}_{1}}^{\boldsymbol{(0)}ba} -
\sdelta\hat{\Sigma}_{\mfrak{I}_{2}\mfrak{I}_{1}}^{ba}  \;.
\end{align}
\end{subequations}
We disregard the detailed model and theory of Refs.\cite{pulse1}-\cite{BCS-SM} with the semiconductor-related solid coupled to ultra-short laser-light pulses and perform
an expansion of the generic sigma-model \(\ovv{Z[\hat{\mscr{J}}]}\) (\ref{x2_11},\ref{x2_13},\ref{x2_14a}) around its saddle point solution 
\(\hat{\Sigma}_{\mfrak{I}_{2}\mfrak{I}_{1}}^{\boldsymbol{(0)}ba}\) of the self-energy by using the corresponding
Green function \(\hat{\mscr{G}}_{\mfrak{I}_{2}\mfrak{I}_{1}}^{\boldsymbol{(0)}ba}\) (\ref{x2_14b},\ref{x2_14c}).  As a result we can sample random numbers
for \(\sdelta\hat{\Sigma}_{\mfrak{I}_{2}\mfrak{I}_{1}}^{ba}\) around the mean values
\(\hat{\Sigma}_{\mfrak{I}_{2}\mfrak{I}_{1}}^{\boldsymbol{(0)}ba}\) for fluctuation properties. The kind of saddle point
approximation and generating function is generic for other kinds of coherent state path integrals as lattice QCD. After
the expansion around the saddle point solution \(\hat{\Sigma}_{\mfrak{I}_{2}\mfrak{I}_{1}}^{\boldsymbol{(0)}ba}\),
we finally acquire the generating function \(\ovv{Z[\hat{\mscr{J}}]}\) (\ref{x2_15a}-\ref{x2_15d}) in a manner which, according to
the description of the 'Neumann-Ulam' Monte-Carlo sampling in section \ref{x12}, contains
the crucial decomposition \(\hat{G}_{ij}=\hat{M}_{ij}^{-1}=(\hat{1}-\hat{m})_{ij}^{-1}\) into unity minus (or plus) a part  
'\(\hat{m}\propto(\im\,\sdelta t/\hbar)\;\sdelta\hat{\Sigma}_{\mfrak{I}_{2}\mfrak{I}_{1}}^{ba}\)' (\ref{x2_15d}) whose eigenvalues tend to zero for
sufficiently small time steps '\(\propto(\im\,\sdelta t/\hbar)\)'. The presence of the saddle point solution in '\(\hat{m}_{j_{1}j_{3}}\)' (\ref{x2_15d}), which multiplies
the fluctuation term '\(\propto(\im\,\sdelta t/\hbar)\;\sdelta\hat{\Sigma}_{\mfrak{I}_{2}\mfrak{I}_{1}}^{ba}\)' in the
determinants, does not alter this scaling  '\(\propto(\im\,\sdelta t/\hbar)\)' to smaller values because the saddle point solution
\(\hat{\Sigma}_{\mfrak{I}_{2}\mfrak{I}_{1}}^{\boldsymbol{(0)}ba}\) as a mean value approaches a fixed, constant value
for physically smaller sizes of  '\(\propto(\im\,\sdelta t/\hbar)\)'. This property can be strictly proven in an analytic manner by
investigation of the saddle point equation (\ref{x2_14b}) so that the terms as
'\(\hat{m}_{j_{1}j_{3}} = 
\hat{\mscr{G}}_{\mfrak{I}_{1}\mfrak{I}_{2}}^{\boldsymbol{(0)}ab}\;
\hat{\mathrm{S}}^{bb}\;q_{2}\;\eta_{2}\;(\im\tfrac{\sdelta t}{\hbar})\sdelta\hat{\Sigma}_{\mfrak{I}_{2}\mfrak{I}_{3}}^{bc}\;
\hat{\mathrm{S}}^{cc}\;q_{3}\;\eta_{3}\)' (\ref{x2_15d}) also advance towards zero values for smaller time steps  '\(\propto(\im\,\sdelta t/\hbar)\)'.
In the following tabulated relations (\ref{x2_15a}-\ref{x2_15d}), we point out the similarity to the described 'Neumann-Ulam' sampling in section \ref{x12}
with the important split \(\hat{M}=\hat{1}-\hat{m}\) (\ref{x1_1a}-\ref{x1_2b}) by under-bracing the determinants \(\wt{\mbox{DET}}\), their corresponding 
'exponential-trace-log'-relations' \(\wt{\mbox{DET}}(matrix)=\exp\{\mbox{Tr}\;\wt{\ln}(matrix)\}\)' and the inversion of  '\(\hat{M}\)'
in terms of  '\(\hat{m}_{j_{1}j_{3}} = 
\hat{\mscr{G}}_{\mfrak{I}_{1}\mfrak{I}_{2}}^{\boldsymbol{(0)}ab}\;
\hat{\mathrm{S}}^{bb}\;q_{2}\;\eta_{2}\;(\im\tfrac{\sdelta t}{\hbar})\sdelta\hat{\Sigma}_{\mfrak{I}_{2}\mfrak{I}_{3}}^{bc}\;
\hat{\mathrm{S}}^{cc}\;q_{3}\;\eta_{3}\)' (\ref{x2_15d}). Note, that the linear term of '\(\hat{m}_{j_{1}j_{3}}\)' (\ref{x2_15d})
has to be cancelled in the expansion of the logarithm, due to the saddle point approximation (\ref{x2_14a}-\ref{x2_14c}) 
which sets linear terms of \(\sdelta\hat{\Sigma}_{\mfrak{I}_{2}\mfrak{I}_{1}}^{ba}\) in the exponents of the derived
generating functions \(\ovv{Z[\hat{\mscr{J}}]}\) (\ref{x2_15a}-\ref{x2_15b}) to zero.  This latter peculiarity is emphasized by
a tilde above the determinants and the logarithms within  the
'exponential-trace-log'-relations' \(\wt{\mbox{DET}}(matrix)=\exp\{\mbox{Tr}\;\wt{\ln}(matrix)\}\)'  (cf. \ref{x2_15c})
\begin{subequations}
\begin{align} \notag &\mbox{\bf Generic case around mean value \(\boldsymbol{\hat{\Sigma}_{\mfrak{I}_{2}\mfrak{I}_{1}}^{\boldsymbol{(0)}ba}}\)
with sampling of \(\boldsymbol{\sdelta\hat{\Sigma}_{\mfrak{I}_{2}\mfrak{I}_{1}}^{ba}}\) :}     \\  \lb{x2_15a} &
\ovv{Z[\hat{\mscr{J}}]} =
\exp\bigg\{\negthickspace-\tfrac{\im}{4}\tfrac{\sdelta t}{\hbar}\sum_{'\!1','\!2'}
(\hat{\mscr{V}}_{\mfrak{i}_{2}\mfrak{i}_{1}})^{-1}
\TRAB\Big[\hat{\mathrm{S}}^{bb}\;q_{2}\;\eta_{2}\;\hat{\Sigma}_{\mfrak{I}_{2}\mfrak{I}_{1}}^{\boldsymbol{(0)}ba}\;
\hat{\mathrm{S}}^{aa}\;q_{1}\;\eta_{1}\;\hat{\Sigma}_{\mfrak{I}_{1}\mfrak{I}_{2}}^{\boldsymbol{(0)}ab}\Big]\negthickspace\bigg\}\;\;
\mbox{DET}\Big(\hat{\mscr{G}}_{\mfrak{I}_{2}\mfrak{I}_{1}}^{\boldsymbol{(0)}ba}
\Big)^{\boldsymbol{-\tfrac{1}{2}}} \\ \notag &\times
\int d[\sdelta\hat{\Sigma}_{\mfrak{I}_{2}\mfrak{I}_{1}}^{ba}]\;
\exp\bigg\{-\tfrac{\im}{4}\tfrac{\sdelta t}{\hbar}\sum_{'\!1','\!2'}(\hat{\mscr{V}}_{\mfrak{i}_{2}\mfrak{i}_{1}})^{-1}
\TRAB\Big[\hat{\mathrm{S}}^{bb}\;q_{2}\;\eta_{2}\;\sdelta\hat{\Sigma}_{\mfrak{I}_{2}\mfrak{I}_{1}}^{ba}\;
\hat{\mathrm{S}}^{aa}\;q_{1}\;\eta_{1}\;\sdelta\hat{\Sigma}_{\mfrak{I}_{1}\mfrak{I}_{2}}^{ab}\Big]\bigg\}\times 
 \\ \notag &\times \underbrace{
\wt{\mbox{DET}}\bigg(\Big[\unityb-\hat{\mscr{G}}_{\mfrak{I}_{1}\mfrak{I}_{2}}^{\boldsymbol{(0)}ab}\;
\hat{\mathrm{S}}^{bb}\;q_{2}\;\eta_{2}\;\big(\im\tfrac{\sdelta t}{\hbar}\big)\,\sdelta\hat{\Sigma}_{\mfrak{I}_{2}\mfrak{I}_{3}}^{bc}\;
\hat{\mathrm{S}}^{cc}\;q_{3}\;\eta_{3}\Big]_{\mfrak{I}_{1}\mfrak{I}_{3}}^{ac}\bigg)^{\boldsymbol{+\tfrac{1}{2}}} 
}_{\scr\big[
\wt{\mbox{\scz DET}}\big(\,\big[\hat{1}-\hat{m}\big]\,\big)^{+\frac{1}{2}}=
\exp\big\{\frac{1}{2}\mbox{\scz Tr}\;\wt{\ln}\big[\hat{1}-\hat{m}\big]\big\} =
\exp\big\{-\frac{1}{2}\;\big(\frac{1}{2}\mbox{\scz Tr}[\,(\hat{m})^{2}\,]+\frac{1}{3}\mbox{\scz Tr}[\,(\hat{m})^{3}\,]+
\frac{1}{4}\mbox{\scz Tr}[\,(\hat{m})^{4}\,]+\ldots\big)\big\}\;\big]}   \times  \\ \notag &\times\;\;
\wt{\mbox{DET}}\bigg(\unityb+\hspace*{-0.42cm}
\underbrace{\Big[\unityb-\hat{\mscr{G}}_{\mfrak{I}_{1}\mfrak{I}_{2}}^{\boldsymbol{(0)}ab}\;
\hat{\mathrm{S}}^{bb}\;q_{2}\;\eta_{2}\;\big(\im\tfrac{\sdelta t}{\hbar}\big)\,\sdelta\hat{\Sigma}_{\mfrak{I}_{2}\mfrak{I}_{3}}^{bc}\;
\hat{\mathrm{S}}^{cc}\;q_{3}\;\eta_{3}\Big]_{\mfrak{I}_{1}\mfrak{I}_{3}}^{\boldsymbol{-1};ac}
 }_{\scr\big[\mbox{\scz(\ref{x1_1a}-\ref{x1_2b}): }
[\hat{1}-\hat{m}]_{ij}^{-1} = \hat{1}_{ij} + \hat{m}_{ij} + \hat{m}_{ii_{2}}\,\hat{m}_{i_{2}j} +
\hat{m}_{ii_{2}}\,\hat{m}_{i_{2}i_{3}}\,\hat{m}_{i_{3}j} + \ldots \big]} \hspace*{-0.36cm}
\hat{\mscr{G}}^{\boldsymbol{(0)}}\;\eta\;\hat{\mscr{J}}\;\eta\bigg)^{\boldsymbol{+\tfrac{1}{2}}} \;; \\      \lb{x2_15b}&
\ovv{Z[\hat{\mscr{J}}]} =
\exp\bigg\{\negthickspace-\tfrac{\im}{4}\tfrac{\sdelta t}{\hbar}\sum_{'\!1','\!2'}
(\hat{\mscr{V}}_{\mfrak{i}_{2}\mfrak{i}_{1}})^{-1}
\TRAB\Big[\hat{\mathrm{S}}^{bb}\;q_{2}\;\eta_{2}\;\hat{\Sigma}_{\mfrak{I}_{2}\mfrak{I}_{1}}^{\boldsymbol{(0)}ba}\;
\hat{\mathrm{S}}^{aa}\;q_{1}\;\eta_{1}\;\hat{\Sigma}_{\mfrak{I}_{1}\mfrak{I}_{2}}^{\boldsymbol{(0)}ab}\Big]\negthickspace\bigg\}\;\;
\mbox{DET}\Big(\hat{\mscr{G}}_{\mfrak{I}_{2}\mfrak{I}_{1}}^{\boldsymbol{(0)}ba}
\Big)^{\boldsymbol{-\tfrac{1}{2}}} \\ \notag &\times
\int d[\sdelta\hat{\Sigma}_{\mfrak{I}_{2}\mfrak{I}_{1}}^{ba}]\;
\exp\bigg\{-\tfrac{\im}{4}\tfrac{\sdelta t}{\hbar}\sum_{'\!1','\!2'}(\hat{\mscr{V}}_{\mfrak{i}_{2}\mfrak{i}_{1}})^{-1}
\TRAB\Big[\hat{\mathrm{S}}^{bb}\;q_{2}\;\eta_{2}\;\sdelta\hat{\Sigma}_{\mfrak{I}_{2}\mfrak{I}_{1}}^{ba}\;
\hat{\mathrm{S}}^{aa}\;q_{1}\;\eta_{1}\;\sdelta\hat{\Sigma}_{\mfrak{I}_{1}\mfrak{I}_{2}}^{ab}\Big]\bigg\}\times 
 \\ \notag &\times \underbrace{
\exp\bigg\{\tfrac{1}{2}\mcal{N}_{x}\sum_{\vec{x}}\sum_{j=0}^{2N+1}
\TRAB\;\wt{\ln}\Big(\Big[\unityb-\hat{\mscr{G}}_{\mfrak{I}_{1}\mfrak{I}_{2}}^{\boldsymbol{(0)}ab}\;
\hat{\mathrm{S}}^{bb}\;q_{2}\;\eta_{2}\;\big(\im\tfrac{\sdelta t}{\hbar}\big)\,\sdelta\hat{\Sigma}_{\mfrak{I}_{2}\mfrak{I}_{3}}^{bc}\;
\hat{\mathrm{S}}^{cc}\;q_{3}\;\eta_{3}\Big]_{\mfrak{I}_{1}\mfrak{I}_{3}}^{ac}\Big) \bigg\}
}_{\scr\big[
\wt{\mbox{\scz DET}}\big(\,\big[\hat{1}-\hat{m}\big]\,\big)^{+\frac{1}{2}}=
\exp\big\{\frac{1}{2}\mbox{\scz Tr}\;\wt{\ln}\big[\hat{1}-\hat{m}\big]\big\} =
\exp\big\{-\frac{1}{2}\;\big(\frac{1}{2}\mbox{\scz Tr}[\,(\hat{m})^{2}\,]+\frac{1}{3}\mbox{\scz Tr}[\,(\hat{m})^{3}\,]+
\frac{1}{4}\mbox{\scz Tr}[\,(\hat{m})^{4}\,]+\ldots\big)\big\}\;\big]}   \times 
  \\ \notag &\times
\exp\bigg\{\tfrac{1}{2}\mcal{N}_{x}\!\!\sum_{\vec{x}}\!\!\sum_{j=0}^{2N+1}\!\TRAB\;\wt{\ln}
\Big(\unityb+\hspace*{-0.42cm}\underbrace{\Big[\unityb-\hat{\mscr{G}}_{\mfrak{I}_{1}\mfrak{I}_{2}}^{\boldsymbol{(0)}ab}\;
\hat{\mathrm{S}}^{bb}\;q_{2}\;\eta_{2}\;\big(\im\tfrac{\sdelta t}{\hbar}\big)\,\sdelta\hat{\Sigma}_{\mfrak{I}_{2}\mfrak{I}_{3}}^{bc}\;
\hat{\mathrm{S}}^{cc}\;q_{3}\;\eta_{3}\Big]_{\mfrak{I}_{1}\mfrak{I}_{3}}^{\boldsymbol{-1};ac}
 }_{\scr\big[\mbox{\scz(\ref{x1_1a}-\ref{x1_2b}): }
[\hat{1}-\hat{m}]_{ij}^{-1} = \hat{1}_{ij} + \hat{m}_{ij} + \hat{m}_{ii_{2}}\,\hat{m}_{i_{2}j} +
\hat{m}_{ii_{2}}\,\hat{m}_{i_{2}i_{3}}\,\hat{m}_{i_{3}j} + \ldots \big]}   \hspace*{-0.36cm}
\hat{\mscr{G}}^{\boldsymbol{(0)}}\;\eta\;\hat{\mscr{J}}\;\eta\Big)^{\boldsymbol{+\tfrac{1}{2}}} 
\hat{\mscr{G}}^{\boldsymbol{(0)}}\;\eta\;\hat{\mscr{J}}\;\eta\Big)\bigg\}_{\!\!\mbox{;}}
\\  \lb{x2_15c} &
\Big(\mbox{\bf\scz linear '\(\boldsymbol{\sdelta\hat{\Sigma}_{\mfrak{I}_{2}\mfrak{I}_{1}}^{ab}}\)'-term has to be cancelled in the exponent of 
'\(\boldsymbol{\wt{\mbox{DET}}(matrix)=\exp\{\TRAB\;\wt{\ln}(matrix)\}}\)'} \Big)\;;  \\ \lb{x2_15d} &\hspace*{1.0cm}
\boldsymbol{\hat{m}_{j_{1}j_{3}} := 
\hat{\mscr{G}}_{\mfrak{I}_{1}\mfrak{I}_{2}}^{\boldsymbol{(0)}ab}\;
\hat{\mathrm{S}}^{bb}\;q_{2}\;\eta_{2}\;\big(\im\tfrac{\sdelta t}{\hbar}\big)\,\sdelta\hat{\Sigma}_{\mfrak{I}_{2}\mfrak{I}_{3}}^{bc}\;
\hat{\mathrm{S}}^{cc}\;q_{3}\;\eta_{3} \;.}
\end{align}
\end{subequations}
The condition on \(\hat{m}_{ij}\) (\ref{x1_1a}-\ref{x1_2b}) in \(\hat{G}_{ij}=\hat{M}_{ij}^{-1}=(\hat{1}-\hat{m})_{ij}^{-1}\), and the restriction of the eigenvalues to be within the open interval \((-1,+1)\), is particularly applicable for the HS-transformed, coherent state path integral \(\ovv{Z[\hat{\mscr{J}}]}\) (\ref{x2_15a}-\ref{x2_15d}) because the analogous self-energy matrix or term in place of \(\hat{m}_{ij}\)  (\ref{x2_15d})
turns out to be proportional to the time step \(\im\,\sdelta t/\hbar\) times the self-energy matrix 
\(\sdelta\hat{\Sigma}_{\mfrak{I}_{2}\mfrak{I}_{1}}^{ba}=
\sdelta\hat{\Sigma}_{\mu_{2},s_{2};\mu_{1},s_{1}}^{ba}(\Teta{j_{2}}\!\!,\vec{x}_{2};\Teta{j_{1}}\!\!,\vec{x}_{1})\) 
itself with incorporation of the Green function factor \(\hat{\mscr{G}}_{\mfrak{I}_{1}\mfrak{I}_{2}}^{\boldsymbol{(0)}ab}\) (\ref{x2_14b}).
The condition on the eigenvalues of \(\hat{m}\propto (\im\,\sdelta t/\hbar)\;\sdelta\hat{\Sigma}\) is thus always achievable
for sufficiently small, time-step intervals \(\sdelta t\) which has anyhow to be the condition for proper, analytical and
numerical considerations. As one is inclined towards smaller time steps with the analogous self-energy terms
 \(\hat{m}\propto (\im\,\sdelta t/\hbar)\;\sdelta\hat{\Sigma}\) for a fixed time \(t_{0},t_{1},\ldots,t_{N}\), \((\sdelta t=t_{N}/N)\)
of observation, one attains smaller eigenvalues of \(\hat{m}\propto (\im\,\sdelta t/\hbar)\;\sdelta\hat{\Sigma}\) and a better
convergence of  the expansion \(\hat{M}_{ij}^{-1}=(\hat{1}-\hat{m})_{ij}^{-1}\), but has to carry out the random sampling
for higher dimensional matrices \(\sdelta\hat{\Sigma}_{\mfrak{I}_{2}\mfrak{I}_{1}}^{ba}=
\sdelta\hat{\Sigma}_{\mu_{2},s_{2};\mu_{1},s_{1}}^{ba}(\Teta{j_{2}}\!\!,\vec{x}_{2};\Teta{j_{1}}\!\!,\vec{x}_{1})\).
However, larger time steps '\(\sdelta t\)' reduce the effort of the sampling of the random, 'fluctuation' self-energy for a fixed
observation time, but may corrupt the condition on the eigenvalues of 
\(\hat{m}\propto (\im\,\sdelta t/\hbar)\;\sdelta\hat{\Sigma}\)
to be within a sufficiently small interval of \((-1,+1)\). Too large values of '\(\sdelta t\)' can even result in a total failure
of the general time development of the considered model. Therefore, a crucial choice in between the two described extremes
of \(\hat{m}\propto (\im\,\sdelta t/\hbar)\;\sdelta\hat{\Sigma}\)  may lead to the most adequate sampling around the
saddle point for a practicable numerical investigation. The condition on a proper value of  '\(\sdelta t\)' 
has to be combined with the expression of the variance (\ref{x1_7a},\ref{x1_10}) 
which includes the matrix \(\hat{K}_{ij}=\hat{m}_{ij}\;\hat{\msf{w}}_{ij}\) (\ref{x1_7b},\ref{x1_10}). Larger transition probabilities \(\hat{\msf{p}}_{ij}\) imply smaller residual weights \(\hat{\msf{w}}_{ij}\) and hence a smaller value
of \((\hat{1}-\hat{K})_{ij}^{-1}\), but increase the inverted stopping probability \(1/\msf{P}\!_{j}\), resulting to an overall
larger variance. Smaller transition probabilities \(\hat{\msf{p}}_{ij}\) generally cause larger residual weights \(\hat{\msf{w}}_{ij}\) and 
therefore a larger value of \((\hat{1}-\hat{K})_{ij}^{-1}\), but decrease the inverted stopping probability \(1/\msf{P}\!_{j}\).
In consequence a good result of the 'Neumann-Ulam' sampling for (\ref{x2_15a}-\ref{x2_15c}) requires a subtle compromise
in the choice \(\hat{m}_{ij}=\hat{\msf{p}}_{ij}\,\hat{\msf{w}}_{ij}\) of the transition probabilities and of the scores with the
residual weights in order to achieve a small variance (\ref{x1_7a}). 
Similar considerations hold for the modified 'Neumann-Ulam' sampling with a partial removal of the stopping probability (\ref{x1_10}).

\section{Conclusions}\lb{x3}

\subsection{Choice of probability densities and approximation schemes of the random sampling}\lb{x31}

An extension of the 'Neumann-Ulam' matrix inversion can be proposed for rather general functions
\([f(\hat{M})]_{ij}=[f(\hat{1}-\hat{m})]_{ij}\) of a matrix \(\hat{M}_{ij}=\hat{1}_{ij}-\hat{m}_{ij}\),
provided that the eigenvalues of \(\hat{m}_{ij}\) are sufficiently placed near zero so that a Taylor expansion
of \([f(\hat{1}-\hat{m})]\) converges with respect to the matrix \(\hat{m}_{ij}\).
Neglecting the first line of \(\ovv{Z[\hat{\mscr{J}}]}\) (\ref{x2_15a}-\ref{x2_15b}) with the exponential of the actions
from the saddle point solution \(\hat{\Sigma}_{\mfrak{I}_{1}\mfrak{I}_{2}}^{\boldsymbol{(0)}ab}\), we can separate
the remaining integrand in the path integrations over \(\sdelta\hat{\Sigma}_{\mfrak{I}_{2}\mfrak{I}_{1}}^{ba}\) (\ref{x3_1a}-\ref{x3_1b})
into the probability density \(\mfrak{R}[\sdelta\hat{\Sigma}_{\mfrak{I}_{2}\mfrak{I}_{1}}^{ba}]\) (\ref{x3_1c}),
accompanied by a phase factor \(\mfrak{P}[\sdelta\hat{\Sigma}_{\mfrak{I}_{2}\mfrak{I}_{1}}^{ba}]\) (\ref{x3_1d}),
times the Green function \(\hat{\mfrak{G}}_{\mfrak{I}_{3}\mfrak{I}_{1}}^{bc}\) (\ref{x3_1e}) which
can be computed according to the 'Neumann-Ulam' matrix inversion because of the given split into \((\hat{1}-\hat{m})_{ij}^{-1}\)
with \(\hat{m}_{j_{1}j_{3}} = 
\hat{\mscr{G}}_{\mfrak{I}_{1}\mfrak{I}_{2}}^{\boldsymbol{(0)}ab}\;
\hat{\mathrm{S}}^{bb}\;q_{2}\;\eta_{2}\;\big(\im\tfrac{\sdelta t}{\hbar}\big)\,\sdelta\hat{\Sigma}_{\mfrak{I}_{2}\mfrak{I}_{3}}^{bc}\;
\hat{\mathrm{S}}^{cc}\;q_{3}\;\eta_{3} \) (\ref{x2_15d},\ref{x3_1f}). In consequence one has to take random numbers corresponding to the distribution
\(d[\sdelta\hat{\Sigma}_{\mfrak{I}_{2}\mfrak{I}_{1}}^{ba}]\;\mfrak{R}[\sdelta\hat{\Sigma}_{\mfrak{I}_{2}\mfrak{I}_{1}}^{ba}]\)
which, itself, can also be tremendously simplified and approximated according to the split '\(\hat{1}-\hat{m}\)'
in the determinant \(\wt{\mbox{DET}}(\,[\hat{1}-\hat{m}]\,)^{+\frac{1}{2}}\)
beginning with a Gaussian factor \(\exp\{-\frac{1}{4}\mbox{Tr}[\,(\hat{m})^{2}\,]\) (cf. (\ref{x2_15c},\ref{x3_1g}))!
It is possible to shift remaining higher order terms of traces to the Green function \(\hat{\mfrak{G}}_{\mfrak{I}_{3}\mfrak{I}_{1}}^{bc}\) (\ref{x3_1e})
which yields a modified observable. Then the choice of transition probabilities and residual weights has to be
determined so that one can sample random walks for the inverted matrix or Green function \(\hat{\mfrak{G}}_{\mfrak{I}_{3}\mfrak{I}_{1}}^{bc}\)
following from the described 'Neumann-Ulam' matrix inversion method of section \ref{x12}
\begin{subequations}
\begin{align} \lb{x3_1a} &
\ovv{Z[\hat{\mscr{J}}]} =\bigg(\mbox{action of saddle point solution }\hat{\Sigma}_{\mfrak{I}_{2}\mfrak{I}_{1}}^{\boldsymbol{(0)}ba}\bigg)\times
\\ \notag &\!\!\!\times\!\!
\int\!\! d[\sdelta\hat{\Sigma}_{\mfrak{I}_{2}\mfrak{I}_{1}}^{ba}]\;\;\mfrak{R}[\sdelta\hat{\Sigma}_{\mfrak{I}_{2}\mfrak{I}_{1}}^{ba}]\;\;
\mfrak{P}[\sdelta\hat{\Sigma}_{\mfrak{I}_{2}\mfrak{I}_{1}}^{ba}]\;\;
\wt{\mbox{DET}}\Big(\unityb+\!\!\Big[\unityb-\hat{\mscr{G}}_{\mfrak{I}_{1}\mfrak{I}_{2}}^{\boldsymbol{(0)}ab}\;
\hat{\mathrm{S}}^{bb}\;q_{2}\;\eta_{2}\;\big(\im\tfrac{\sdelta t}{\hbar}\big)\,\sdelta\hat{\Sigma}_{\mfrak{I}_{2}\mfrak{I}_{3}}^{bc}\;
\hat{\mathrm{S}}^{cc}\;q_{3}\;\eta_{3}\Big]_{\mfrak{I}_{1}\mfrak{I}_{3}}^{\boldsymbol{-1};ac}\,
\hat{\mscr{G}}^{\boldsymbol{(0)}}\;\eta\;\hat{\mscr{J}}\;\eta\Big)^{\!\boldsymbol{+\tfrac{1}{2}}}_{\mbox{;}}  \\  \lb{x3_1b} &\!\!\!
\frac{\pp\ovv{Z[\hat{\mscr{J}}]}}{\pp\hat{\mscr{J}}_{\mfrak{I}_{4}\mfrak{I}_{1}}^{da}} \!=\! 
\Big\langle\!\hat{\mfrak{G}}_{\mfrak{I}_{1}\mfrak{I}_{4}}^{ad} \Big\rangle
\!=\!\!\bigg(\mbox{action of saddle point solution }\hat{\Sigma}_{\mfrak{I}_{2}\mfrak{I}_{1}}^{\boldsymbol{(0)}ba}\bigg)\!\!\times\!\!
\int\!\! d[\sdelta\hat{\Sigma}_{\mfrak{I}_{2}\mfrak{I}_{1}}^{ba}]\;\;\mfrak{R}[\sdelta\hat{\Sigma}_{\mfrak{I}_{2}\mfrak{I}_{1}}^{ba}]\;\;
\mfrak{P}[\sdelta\hat{\Sigma}_{\mfrak{I}_{2}\mfrak{I}_{1}}^{ba}]\;\;\hat{\mfrak{G}}_{\mfrak{I}_{1}\mfrak{I}_{4}}^{ad}\,;   \\  \lb{x3_1c}    &\mfrak{R}[\sdelta\hat{\Sigma}_{\mfrak{I}_{2}\mfrak{I}_{1}}^{ba}]=     
\exp\bigg\{-\tfrac{1}{4}\tfrac{\sdelta t}{\hbar}\sum_{'\!1','\!2'}\mathsf{Real}[\,\im\;(\hat{\mscr{V}}_{\mfrak{i}_{2}\mfrak{i}_{1}})^{-1}]\;\;
\TRAB\Big[\hat{\mathrm{S}}^{bb}\;q_{2}\;\eta_{2}\;\sdelta\hat{\Sigma}_{\mfrak{I}_{2}\mfrak{I}_{1}}^{ba}\;
\hat{\mathrm{S}}^{aa}\;q_{1}\;\eta_{1}\;\sdelta\hat{\Sigma}_{\mfrak{I}_{1}\mfrak{I}_{2}}^{ab}\Big]\bigg\}\times 
\\ \notag &\times 
\exp\bigg\{\tfrac{1}{2}\;\mathsf{Real}\Big[\mcal{N}_{x}{\ts\sum_{\vec{x}}\sum_{j=0}^{2N+1}}
\TRAB\;\wt{\ln}\Big(\Big[\unityb-\hat{\mscr{G}}_{\mfrak{I}_{1}\mfrak{I}_{2}}^{\boldsymbol{(0)}ab}\;
\hat{\mathrm{S}}^{bb}\;q_{2}\;\eta_{2}\;\big(\im\tfrac{\sdelta t}{\hbar}\big)\,\sdelta\hat{\Sigma}_{\mfrak{I}_{2}\mfrak{I}_{3}}^{bc}\;
\hat{\mathrm{S}}^{cc}\;q_{3}\;\eta_{3}\Big]_{\mfrak{I}_{1}\mfrak{I}_{3}}^{ac}\Big) \Big]\bigg\}\;;    \\       \lb{x3_1d}   
&\mfrak{P}[\sdelta\hat{\Sigma}_{\mfrak{I}_{2}\mfrak{I}_{1}}^{ba}]=     
\exp\bigg\{-\tfrac{\im}{4}\tfrac{\sdelta t}{\hbar}\sum_{'\!1','\!2'}\mathsf{Imag}[\,\im\;(\hat{\mscr{V}}_{\mfrak{i}_{2}\mfrak{i}_{1}})^{-1}]\;
\TRAB\Big[\hat{\mathrm{S}}^{bb}\;q_{2}\;\eta_{2}\;\sdelta\hat{\Sigma}_{\mfrak{I}_{2}\mfrak{I}_{1}}^{ba}\;
\hat{\mathrm{S}}^{aa}\;q_{1}\;\eta_{1}\;\sdelta\hat{\Sigma}_{\mfrak{I}_{1}\mfrak{I}_{2}}^{ab}\Big]\bigg\}\times 
\\ \notag &\times 
\exp\bigg\{\tfrac{\im}{2}\;\mathsf{Imag}\Big[\mcal{N}_{x}{\ts\sum_{\vec{x}}\sum_{j=0}^{2N+1}}
\TRAB\;\wt{\ln}\Big(\Big[\unityb-\hat{\mscr{G}}_{\mfrak{I}_{1}\mfrak{I}_{2}}^{\boldsymbol{(0)}ab}\;
\hat{\mathrm{S}}^{bb}\;q_{2}\;\eta_{2}\;\big(\im\tfrac{\sdelta t}{\hbar}\big)\,\sdelta\hat{\Sigma}_{\mfrak{I}_{2}\mfrak{I}_{3}}^{bc}\;
\hat{\mathrm{S}}^{cc}\;q_{3}\;\eta_{3}\Big]_{\mfrak{I}_{1}\mfrak{I}_{3}}^{ac}\Big)\Big] \bigg\}\;;   \\          \lb{x3_1e}   & 
\hat{\mfrak{G}}_{\mfrak{I}_{1}\mfrak{I}_{4}}^{ad} =\eta_{1}     
\Big[\unityb-\hat{\mscr{G}}_{\mfrak{I}_{1}\mfrak{I}_{2}}^{\boldsymbol{(0)}ab}\;
\hat{\mathrm{S}}^{bb}\;q_{2}\;\eta_{2}\;\big(\im\tfrac{\sdelta t}{\hbar}\big)\,\sdelta\hat{\Sigma}_{\mfrak{I}_{2}\mfrak{I}_{3}}^{bc}\;
\hat{\mathrm{S}}^{cc}\;q_{3}\;\eta_{3}\Big]_{\mfrak{I}_{1}\mfrak{I}_{3}}^{\boldsymbol{-1};ac}\;     
\hat{\mscr{G}}_{\mfrak{I}_{3}\mfrak{I}_{4}}^{\boldsymbol{(0)}cd}\;\eta_{4} =
\eta_{1}\;\big[\hat{1}-\hat{m}\big]_{\mfrak{I}_{1}\mfrak{I}_{3}}^{\boldsymbol{-1};ac}\; 
\hat{\mscr{G}}_{\mfrak{I}_{3}\mfrak{I}_{4}}^{\boldsymbol{(0)}cd}\;\eta_{4}\;;
\\  \lb{x3_1f}  &    \hat{m}_{j_{1}j_{3}} = 
\hat{\mscr{G}}_{\mfrak{I}_{1}\mfrak{I}_{2}}^{\boldsymbol{(0)}ab}\;
\hat{\mathrm{S}}^{bb}\;q_{2}\;\eta_{2}\;\big(\im\tfrac{\sdelta t}{\hbar}\big)\,\sdelta\hat{\Sigma}_{\mfrak{I}_{2}\mfrak{I}_{3}}^{bc}\;
\hat{\mathrm{S}}^{cc}\;q_{3}\;\eta_{3} \;;     \\     \lb{x3_1g}                  &\wt{\mbox{DET}}(\,[\hat{1}-\hat{m}]\,)^{\boldsymbol{+\frac{1}{2}}}=
\exp\Big\{\frac{1}{2}\mbox{Tr}\;\wt{\ln}\big[\hat{1}-\hat{m}\big]\Big\} =
\exp\Big\{\!\!-\frac{1}{2}\Big(\frac{1}{2}\mbox{Tr}[\,(\hat{m})^{2}\,]+\!\frac{1}{3}\mbox{Tr}[\,(\hat{m})^{3}\,]+\!
\frac{1}{4}\mbox{Tr}[\,(\hat{m})^{4}\,]+\ldots\!\Big)\!\Big\}_{\mbox{.}}
\end{align}
\end{subequations}
A possible choice of transition probabilities  \(\hat{\msf{p}}_{ij}\) and residual weights \(\hat{\msf{w}}_{ij}\) for a sparse matrix \(\hat{m}_{ij} = \hat{\msf{p}}_{ij}\,\hat{\msf{w}}_{ij}\) (\ref{x3_1f})
should result from the factorization 
\begin{align}\lb{x3_2}
\hat{\msf{p}}_{ij} &=s_{0}\;\sqrt{|\hat{m}_{ij}|};\;\;\;|\hat{\msf{w}}_{ij}|=\sqrt{|\hat{m}_{ij}|}/s_{0};\;\;\;
\hat{m}_{ij}/|\hat{m}_{ij}|=\hat{\msf{w}}_{ij}/|\hat{\msf{w}}_{ij}|;\;\;\;\hat{m}_{ij},\,\hat{\msf{w}}_{ij}\in\mathbb{C}\;,
\end{align}
with an appropriate scale factor \(s_{0}\) so that a direct numerical inversion can be circumvented. A good choice of stopping and
continuation probabilities \(\msf{P}\!_{j}\), \(\ovv{\msf{P}}\!_{j}=1-\msf{P}\!_{j}\) should be attained for equal 
values of \(\tfrac{1}{2}\) from the relation
\(\msf{P}\!_{j}+\sum_{i}\hat{\msf{p}}_{ij}=\msf{P}\!_{j}+s_{0}\;\sum_{i}\sqrt{|\hat{m}_{ij}|}=1\)
with \(\hat{m}_{ij}\) (\ref{x3_1f}) being proportional to the time interval \(\sdelta t\) and to the saddle point fluctuation
\(\sdelta\hat{\Sigma}_{\mfrak{I}_{2}\mfrak{I}_{1}}^{ba}\). A final adaptation with the overall scale factor \(s_{0}\) determines
the transition probabilities and residual weights. A further choice of transition probabilities can include the physical intuition where
higher precedence is given to those states '\(\mfrak{I}_{1}\)', '\(\mfrak{I}_{2}\)' of the transition probabilities 
'\(\hat{\msf{p}}_{\mfrak{I}_{1}\mfrak{I}_{2}}\)' and residual weights '\(\hat{\msf{w}}_{\mfrak{I}_{1}\mfrak{I}_{2}}\)'
which yield a higher occupation probability of particles in the physically important parts of the spacetime coordinates.

Nevertheless, there is a further problem of the random sampling according to the probability density
\(d[\sdelta\hat{\Sigma}_{\mfrak{I}_{2}\mfrak{I}_{1}}^{ba}]\;\mfrak{R}[\sdelta\hat{\Sigma}_{\mfrak{I}_{2}\mfrak{I}_{1}}^{ba}]\) (\ref{x3_1c}) which, however, occurs in any
Monte-Carlo method with '{\it many}' independent random variables. If one reduces the probability density
\(d[\sdelta\hat{\Sigma}_{\mfrak{I}_{2}\mfrak{I}_{1}}^{ba}]\;\mfrak{R}[\sdelta\hat{\Sigma}_{\mfrak{I}_{2}\mfrak{I}_{1}}^{ba}]\) (\ref{x3_1c}) to the analogous problem of  '\(N_{z}\)' independent, random Gaussian factors
\begin{align}\lb{x3_3}
d[\sdelta\hat{\Sigma}_{\mfrak{I}_{2}\mfrak{I}_{1}}^{ba}]\;\mfrak{R}[\sdelta\hat{\Sigma}_{\mfrak{I}_{2}\mfrak{I}_{1}}^{ba}]&\rightarrow
\prod_{i=1}^{N_{z}}\tfrac{dz_{i}^{*}\;dz_{i}}{2\im}\;\exp\big\{-\tfrac{1}{2}\;z_{i}^{*}\,z_{i}\big/\sigma_{i}^{2}\big\}\;;\;(z_{i}\in\mathbb{C})\;,
\end{align}
the total probability of the  '{\it many}' random Gaussian factors immediately gives a severe loss of a total '{\it relevant hit}'
in phase space with a probability sufficiently close to one. Therefore, it can be advantageous in most cases to change to spherical coordinates
after a rescaling of the individual random variables \(z_{i}\rightarrow z_{i}\,\sigma_{i}\). One thus achieves a probability
distribution of a radial variable \(\rho\) times a surface area \(d\Omega_{2N_{z}-1}\) in the created \(2N_{z}\)-dim.\
Euclidean phase space (with modified scores due to the rescaling with the variances \(\sigma_{i}\))
\begin{align}\lb{x3_4}
\prod_{i=1}^{N_{z}}\tfrac{dz_{i}^{*}\;dz_{i}}{2\im}\;\exp\big\{-\tfrac{1}{2}\;z_{i}^{*}\,z_{i}\big/\sigma_{i}^{2}\big\}\longrightarrow
\Big(z_{i}\rightarrow z_{i}\,\sigma_{i}\Big)\longrightarrow\Big(\prod_{i=1}^{N_{z}}\sigma_{i}^{2}\Big)\;
d\Omega_{2N_{z}-1}\;d\rho\;\rho^{2N_{z}-1}\;\exp\{-\tfrac{1}{2}\;\rho^{2}\big\}\;.
\end{align}
If the corresponding score function only depends weakly on the surface term \(d\Omega_{2N_{z}-1}\), e.g.\ with lowest order
spherical harmonics depending only on a '{\it few}' angular-coordinates, one has obtained an
improvement according to the principle of '{\it importance sampling}', however even, in the case of a many body physics.

Further approximation schemes concern the reduction of our chosen, strongly time dependent problem of ultra-short laser propagation
to the case of a stationary self-energy matrix which is then only subject to the difference \(t_{j_{2}}-t_{j_{1}}\) of its two
contour time arguments \(\Teta{j_{2}}\), \(\Teta{j_{1}}\) (\ref{x2_2}-\ref{x2_4b}) with further combination of the corresponding time contour branches \(\eta_{j_{2}},\,\eta_{j_{1}}=\pm\).
A Fourier transformation then simplifies the whole problem to the case of a self-energy matrix
which only entails an individual frequency variable \(\hat{\Sigma}_{\mu_{2},s_{2};\mu_{1},s_{1}}^{ba}(\Teta{j_{2}}\!,\vec{x}_{2};\Teta{j_{1}}\!,\vec{x}_{1})\rightarrow
\hat{\Sigma}_{\mu_{2},s_{2};\mu_{1},s_{1}}^{ba;\eta_{j_{2}},\eta_{j_{1}}}(\omega;\vec{x}_{2},\vec{x}_{1})\).
This simplifies the solution of the saddle point equation (\ref{x2_14b}) if the physical system is reduced to the case of a stationary time
problem. Other approximations and changes of the random variables \(\sdelta\hat{\Sigma}_{\mfrak{I}_{2}\mfrak{I}_{1}}^{ba}\) 
with modification of the probability density \(\mfrak{R}[\sdelta\hat{\Sigma}_{\mfrak{I}_{2}\mfrak{I}_{1}}^{ba}]\) (\ref{x3_1c}), 
containing the determinant \(\wt{\mbox{DET}}(\,[\hat{1}-\hat{m}]\,)^{+\frac{1}{2}}\), have similarly to be resolved from
the guiding principle of '{\it importance sampling}' in order to acquire a sufficient number of  '{\it relevant hits}'
from the random sampling even in the case of a many body problem.

\end{document}